# Rheology-controlled hydraulic selection in fracture–matrix heat transport: Mechanisms and thermal signatures

Alessandro Lenci [a,b,c,*], Irene Daprà [a]

[a] Department of Civil, Chemical, Environmental, and Materials Engineering (DICAM), University of Bologna, Viale del Risorgimento 2, 40136, Bologna, Italy
[b] Department of Energy Science & Engineering, Stanford University, Green Earth Sciences Building, 367 Panama Street, 94305, Stanford, CA, USA
[c] Univ Rennes, CNRS, Géosciences Rennes, UMR 6118, Campus de Beaulieu, Bâtiment 15, 263 avenue du Général Leclerc, 35042, Rennes Cedex, France

ABSTRACT

Geological fractures exhibit heterogeneous aperture fields that localize flow along preferential pathways and produce nonuniform fluid–matrix contact times. Heat transport combines channelized advection with conductive matrix exchange. For non-Newtonian fluids this coupling is constitutive: shear thinning biases the aperture-to-flux mapping toward larger apertures, while yield stress suppresses flow below the mobilization threshold. Thermal-front advance, longitudinal spreading, and outlet breakthrough diagnose this selection. A stochastic semi-analytical model represents aperture classes as parallel pathways with constitutive fluxes and superposes their channel-scale advection–conduction solutions for a semi-infinite matrix. Matrix diffusion sets the late-time scalings of breakthrough curves and front moments, while aperture variability and rheology control their amplitudes and crossover. Rheology acts on spreading not only through the mean velocity but through high-order flux-weighted aperture moments that set the amplitude of persistent inter-channel variance, making longitudinal variance the most sensitive diagnostic. For monomial flux laws the reduced late-time amplitudes collapse onto a single similarity variable, so shear thinning acts as amplified effective aperture variability. Global sensitivity analysis attributes flux reweighting to shear thinning and hydraulic accessibility to yield stress. Two-dimensional simulations verify the construction for Newtonian and power-law flow; a connected rough-field comparison at low aperture variability recovers mean advance and outlet survival but overestimates variance, so Ellis and Herschel–Bulkley results remain independent-channel predictions. The model is an interpretable reference limit separating rheology-controlled hydraulic selection from matrix-controlled transport: it predicts the breakthrough signature of a rheologically selective thermal tracer during engineered injection, rather than guiding the choice of a long-term production fluid.

## 1. Introduction

Heat transport in fractured geological media is governed by advective flow through permeable fractures and conductive exchange with the surrounding rock matrix. This process is central to geothermal energy extraction, subsurface thermal-energy storage, and heat-tracer characterization of fractured formations (Gringarten et al., 1975; Silliman, 1989; Klepikova et al., 2011; Shaik et al., 2011). Because fracture permeability typically exceeds matrix permeability by orders of magnitude, flow is concentrated within the fracture network, while the matrix provides a large thermal-storage volume (Bonnet et al., 2001; Berkowitz, 2002; Viswanathan et al., 2022). Conductive exchange with the matrix stores heat over progressively increasing penetration depths, so that the thermal history at the fracture–matrix interface continues to affect the fracture response at later times, delaying thermal recovery and generating long-time tails in the breakthrough response (Vinsome and Westerveld, 1980; Neuville et al., 2013; Cherubini et al., 2017). In heterogeneous fractures, aperture variability makes the flow field nonuniform; as a result, the measured thermal response preferentially samples high-conductance pathways with distinct fluid–matrix contact times. The macroscopic thermal response therefore reflects the coupled effects of aperture-controlled velocity heterogeneity and fracture–matrix exchange, which together shape thermal-front spreading, breakthrough broadening, and long-time recovery (Roubinet et al., 2012; Klepikova et al., 2021; Wang et al., 2024; Lenci et al., 2026b).

At the fracture scale, preferential channeling arises because aperture heterogeneity creates spatial contrasts in hydraulic resistance, so that flow partitions into lower-resistance, higher-conductance pathways (Tsang and Tsang, 1987; Moreno et al., 1988). For Newtonian creeping flow between locally parallel walls, slit-flow transmissivity scales with the third power of aperture (Witherspoon et al., 1980; Brown, 1987); even moderate aperture fluctuations can therefore generate large velocity contrasts (Méheust and Schmittbuhl, 2001, 2003). This resistance-controlled flow partitioning is represented, within the lubrication approximation, by depth-averaged descriptions of rough-walled fractures (Brown, 1987; Brush and Thomson, 2003; Krishna et al., 2025). The associated velocity heterogeneity accelerates transport along high-velocity pathways while delaying it in low-velocity regions, thereby enhancing longitudinal spreading and broadening breakthrough responses (Thompson, 1991; Roux et al., 1998; Plouraboué et al., 1998; Drazer et al., 2004). These velocity contrasts are a major source of the non-Fickian signatures observed in solute and heat transport through heterogeneous fractures (Wang and Cardenas, 2014; Klepikova et al., 2021; Lenci et al., 2026a,b).

* Corresponding author at: Department of Civil, Chemical, Environmental, and Materials Engineering (DICAM), University of Bologna, Viale del Risorgimento 2, 40136, Bologna, Italy.
*E-mail address:* alessandro.lenci@unibo.it (A. Lenci).

For a parallel-plate fracture embedded in a semi-infinite matrix, the classical single-fracture advection–conduction solution of Lauwerier (1955) represents longitudinal advection in the fracture coupled to transverse conductive diffusion in the matrix. In reduced fracture–matrix transport models, this coupling appears as a semi-infinite conductive-memory kernel, which sets the diffusive long-time tail of the thermal response. Time-domain random-walk (TDRW) simulations for Newtonian heat transport in rough-walled fractures have shown that the same kernel can be represented by a one-sided stable trapping-time law (De Simone et al., 2021; Lenci et al., 2026b). Extensions to heterogeneous fractures and fracture networks have shown that aperture variability and flow channeling control early-to-intermediate thermal breakthrough, thermal tailing, and fracture–matrix exchange efficiency (Neuville et al., 2013; Cherubini et al., 2017; De Simone et al., 2021; Dejam and Hassanzadeh, 2023). Field heat-tracing experiments with hot and cold water in fractured aquifers report the same matrix-diffusion signature and reproduce it with process-based fracture–matrix transport models (Hoffmann et al., 2022). Aperture heterogeneity then controls how the asymptotic regime is approached, through flux-weighted advective time scales, tail amplitudes, and the crossover from aperture-controlled early transport to matrix-controlled late-time behavior (Hyman et al., 2019).

Engineered non-Newtonian fluids are widely used for mobility and sweep control and as viscosified carriers (Sorbie, 1991; Tosco et al., 2014; Paria, 2008). Their stress-dependent flow partitioning also makes them diagnostic: polymer tracers reveal weak permeability heterogeneity (D'Onofrio et al., 2002), and non-Newtonian injection has been used to infer hydraulic conductivity, pore-size distributions, and rough-fracture apertures (Abou Najm and Atallah, 2016; Rodríguez de Castro et al., 2020). Related heat-transfer applications include geothermal drilling fluids (Santoyo et al., 2003), nanoparticle-enhanced geothermal fluids (Soltani et al., 2022), and elastic-instability-driven mixing in porous media (Browne and Datta, 2024); most fracture–matrix heat-transport models nevertheless remain Newtonian.

Water and supercritical $CO_2$ remain the standard fluids for long-term thermal production; non-Newtonian formulations are used mainly for fracture sustainment, mobility control, and carrier placement (Barbati et al., 2016). This study instead targets thermal characterization during engineered injection. A controlled temperature perturbation carried by the injected fluid acts as a rheologically selective tracer: the constitutive law determines which aperture classes carry the signal, while breakthrough records their contact-time exchange with the matrix. Because this selection cannot generally be represented by a single effective viscosity (Roustaei et al., 2016), the relevant in-situ rheology must be known over the injection and observation window.

Rheology–geometry coupling has been documented in power-law backflow, gravity currents, drainage, and dispersion (Lenci et al., 2021; Longo et al., 2021; Zeighami et al., 2022; Chiapponi et al., 2020). In fractures, shear thinning can enhance channelization (Auradou et al., 2008; Lavrov, 2013; Rodríguez de Castro and Radilla, 2016; Lenci et al., 2022a, 2024), whereas a yield stress can exclude apertures below the activation threshold (Boronin et al., 2015; Rodríguez de Castro et al., 2020). Heat transport records this hydraulic selection through flux weights, advective time scales, and matrix exchange.

We develop a semi-analytical aperture-distributed model based on reduced channel representations (Neuzil and Tracy, 1981; Tsang and Tsang, 1987) and validated non-Newtonian conductance laws (Felisa et al., 2018; Lenci and Di Federico, 2020; Lenci et al., 2020; Lenci and Chiapponi, 2020). Flux-weighted superposition of single-channel Lauwerier solutions introduces shear-thinning reweighting and yield-threshold activation directly. A variance-based GSA identifies their dominant controls, while two-dimensional TDRW simulations (Lenci et al., 2026b) verify the construction in an exact parallel-channel geometry and assess one connected rough-field realization over aperture coefficients of variation $0.05 \le \mathrm{CV}_a \le 0.15$. The latter comparison does not validate the Ellis or Herschel–Bulkley predictions in connected fields. The analysis also derives a one-parameter similarity collapse for monomial flux laws, an asymptotically slow yield-edge correction, and a weak-heterogeneity screening reference for persistent inter-channel spreading.

Sections 2–5 develop the hydraulic closures, transport model, analytical limits, and similarity structure; supporting algebra is collected in Appendix D. Section 6 defines the numerical and GSA settings, Section 7 presents the results and numerical assessment, and Section 8 concludes.

## 2. Hydraulic response of a single fracture channel

The hydraulic response of an individual fracture channel is defined for an idealized parallel-plate slit of aperture $a$ and unit width. A Cartesian coordinate system is attached to the channel, with $x_1$ denoting the longitudinal coordinate along the channel axis, $x_2$ the transverse coordinate across the unit width, and $x_3$ the wall-normal coordinate. The channel mid-plane is located at $x_3 = 0$, and the two walls are located at $x_3 = \pm a/2$ (Fig. 1). The channel has smooth no-slip walls and uniform aperture along its length, and is driven by a prescribed constant longitudinal pressure-gradient magnitude $|\partial_{x_1}\mathcal{P}| = \Delta\mathcal{P}/L$, where $\Delta\mathcal{P}$ is the pressure drop over the channel length $L$. The flow is assumed to be steady, incompressible, inertia-free, fully developed, and unidirectional. The velocity field is therefore written as $\mathbf{u} = u(x_3)\mathbf{i}_1$. For such uniform channels the advective acceleration vanishes identically, so the unidirectional profile solves the full momentum balance at any laminar Reynolds number; inertial and roughness-induced deviations arising in real fractures (Cardenas et al., 2009) lie outside this reference description.

The longitudinal momentum balance for the shear stress $\tau$ reduces to

$$\frac{\mathrm{d}\tau}{\mathrm{d}x_3} = \partial_{x_1}\mathcal{P}. \tag{2.1}$$

For flow in the positive $x_1$ direction, $\partial_{x_1}\mathcal{P} = -|\partial_{x_1}\mathcal{P}|$. By symmetry, the shear stress vanishes at the centerline, $\tau(0) = 0$, and its absolute value increases linearly toward the walls:

$$|\tau(x_3)| = |\partial_{x_1}\mathcal{P}|\,|x_3|, \tag{2.2a}$$

$$\tau_w(a) = \frac{a}{2}|\partial_{x_1}\mathcal{P}| = \frac{a\Delta\mathcal{P}}{2L}, \tag{2.2b}$$

where $\tau_w$ is the wall shear-stress magnitude.

For the generalized Newtonian closures considered below, this stress distribution is closed by a constitutive relation between the shear-stress magnitude $|\tau|$ and the shear rate

$$\dot\gamma = \left|\frac{\mathrm{d}u}{\mathrm{d}x_3}\right|, \tag{2.3}$$

where $u(x_3)$ is the longitudinal velocity component.

The no-slip boundary conditions are

$$u(-a/2) = u(a/2) = 0. \tag{2.4}$$

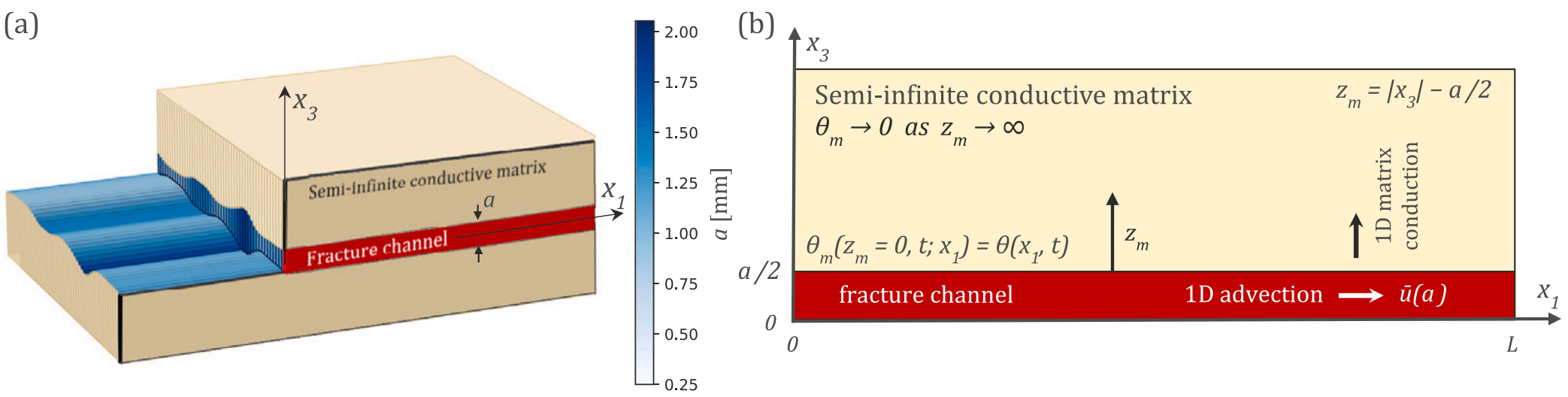

**Fig. 1.** Model geometry. (a) Parallel-channel representation of a variable-aperture fracture: one realization of the aperture set, lognormally distributed with $CV_a = 0.40$, correlated across the transverse direction, and uniform along $x_1$. The upper wall is cut away over the near half of the block, so that the aperture set is seen from above with aperture rendered as color; where the block is intact the fracture is seen edge-on, and one channel of aperture $a$ is indicated. (b) Cross-section of a single channel: longitudinal advection at the aperture-averaged velocity $\bar{u}(a)$ over $0 \le x_1 \le L$, and one-dimensional conduction in the semi-infinite matrix along the wall-normal coordinate $z_m = |x_3| - a/2$, with continuity of temperature at the fluid–matrix interface. The upper half is shown: the configuration is symmetric about the channel mid-plane $x_3 = 0$, and heat is exchanged through both walls.

Once the constitutive relation is specified, the velocity profile $u(x_3; a)$ follows; the volumetric flux per unit width and the aperture-averaged velocity are then defined by

$$q(a) = \int_{-a/2}^{a/2} u(x_3; a)\, \mathrm{d}x_3, \qquad \bar{u}(a) = \frac{q(a)}{a}. \tag{2.5}$$

Integrating $q(a)$ by parts and changing variable from $x_3$ to the shear stress gives $q(a) = 2|\partial_{x_1}\mathcal{P}|^{-2}\int_0^{\tau_w} \tau\dot{\gamma}(\tau)\,\mathrm{d}\tau$, a functional that is linear in $\dot{\gamma}$. The closures below follow from it directly: the Ellis flux from the additivity of Eq. (2.14), and the Herschel–Bulkley flux from the substitution $\tau \mapsto \tau - \tau_y$.

### 2.1. Newtonian fluid

For a Newtonian fluid, the signed shear stress satisfies

$$\tau = \eta \frac{\mathrm{d}u}{\mathrm{d}x_3}, \tag{2.6}$$

where $\eta$ is the dynamic viscosity. Combining this constitutive relation with Eq. (2.1) gives

$$\eta \frac{\mathrm{d}^2 u}{\mathrm{d}x_3^2} = \partial_{x_1}\mathcal{P}. \tag{2.7}$$

Integration with the no-slip boundary conditions at $x_3 = \pm a/2$ gives the parabolic velocity profile

$$u(x_3) = \frac{|\partial_{x_1}\mathcal{P}|}{2\eta}\left(\frac{a^2}{4} - x_3^2\right). \tag{2.8}$$

Using the definitions of $q(a)$ and $\bar{u}(a)$ above, the Newtonian flux per unit width and aperture-averaged velocity are

$$q_N(a) = C_N a^3, \qquad \bar{u}_N(a) = C_N a^2, \tag{2.9}$$

with

$$C_N = \frac{|\partial_{x_1}\mathcal{P}|}{12\eta} = \frac{\Delta\mathcal{P}}{12\eta L}. \tag{2.10}$$

The Newtonian channel flux therefore scales cubically with aperture, recovering the local cubic law.

### 2.2. Power-law fluid

A power-law fluid satisfies

$$|\tau| = m\dot{\gamma}^n, \tag{2.11}$$

where $m$ is the consistency index and $n$ is the flow index. Using the stress distribution Eq. (2.2a),

$$q_{PL}(a) = C_{PL} a^{2+1/n}, \qquad \bar{u}_{PL}(a) = C_{PL} a^{1+1/n}, \tag{2.12}$$

with

$$C_{PL} = \frac{n}{2n+1}\left(\frac{1}{2}\right)^{1+1/n}\left(\frac{|\partial_{x_1}\mathcal{P}|}{m}\right)^{1/n}. \tag{2.13}$$

For shear-thinning fluids, $n < 1$, and therefore $2 + 1/n > 3$; the channel flux is more sensitive to aperture than in the Newtonian case.

### 2.3. Ellis fluid

The Ellis model has previously been used to represent bounded-viscosity, shear-thinning fracture flow (Ciriello et al., 2021). Its constitutive law can be written in terms of the stress-dependent apparent viscosity as

$$\dot{\gamma}(\tau) = \frac{|\tau|}{\eta_0}\left[1 + \left(\frac{\eta_0^n |\tau|^{1-n}}{m}\right)^{1/n}\right]. \tag{2.14}$$

Here $\eta_0$ is the zero-shear viscosity, while $m$ and $n$, introduced in Section 2.2, characterize the asymptotic power-law branch.

For an Ellis fluid, the slit flux is

$$q_E(a) = C_N a^3 + C_{PL} a^{2+1/n}. \tag{2.15}$$

Here $q_E(a)$ is the Ellis-fluid flux per unit width, with $C_N$ evaluated from Eq. (2.10) at $\eta = \eta_0$.

The crossover aperture $a_c$ is defined by equality of the two flux contributions,

$$C_N a_c^3 = C_{PL} a_c^{2+1/n}. \tag{2.16}$$

For $n < 1$, this gives

$$a_c = \left(\frac{C_N}{C_{PL}}\right)^{n/(1-n)}. \tag{2.17}$$

For $a \ll a_c$, the Newtonian branch dominates; for $a \gg a_c$, the shear-thinning branch dominates. Equivalently, at any reference aperture $a_0$ the balance between the two branches is measured by the dimensionless branch ratio

$$\chi_E = \frac{C_{PL} a_0^{2+1/n}}{C_N a_0^3} = \left(\frac{a_0}{a_c}\right)^{1/n-1}, \tag{2.18}$$

which is used below as the shape coordinate of the Ellis closure, evaluated at the mean aperture. The mapping between the parameterization used here and the standard Ellis constitutive parameters $(\eta_0, \tau_{1/2}, n)$, in which $\tau_{1/2}$ is the transition stress at which the apparent viscosity has fallen to half its zero-shear value, is given in Appendix A.

### 2.4. Herschel–Bulkley fluid

A Herschel–Bulkley fluid satisfies

$$|\tau| = \tau_y + m\dot{\gamma}^n, \qquad |\tau| \ge \tau_y, \tag{2.19}$$

where $\tau_y$ is the yield stress, and $m$ and $n$ are the consistency and flow indices introduced in Section 2.2. Below yield, $\dot{\gamma} = 0$. Thus, unlike the power-law and Ellis fluid closures, the Herschel–Bulkley law introduces a threshold for hydraulic activation.

Using the wall shear-stress magnitude in Eq. (2.2b), flow occurs only when $\tau_w(a) > \tau_y$. This condition defines the yield aperture

$$a_y = \frac{2\tau_y}{|\partial_{x_1}\mathcal{P}|}. \tag{2.20}$$

Therefore, $q_{HB}(a) = 0$ for $a \le a_y$, whereas for $a > a_y$ the flux per unit width is

$$q_{HB}(a) = \frac{2}{|\partial_{x_1}\mathcal{P}|^2 m^{1/n}}\left[\frac{\left(\tau_w(a)-\tau_y\right)^{2+1/n}}{2+1/n} + \frac{\tau_y\left(\tau_w(a)-\tau_y\right)^{1+1/n}}{1+1/n}\right]. \tag{2.21}$$

Equivalently, introducing

$$\xi(a) = \frac{\tau_y}{\tau_w(a)} = \frac{a_y}{a}, \tag{2.22}$$

the active-aperture flux can be written as

$$q_{HB}(a) = q_{PL}(a)\mathcal{F}_Y(\xi), \qquad a > a_y, \tag{2.23}$$

where

$$\mathcal{F}_Y(\xi) = (1-\xi)^{2+1/n} + \frac{2n+1}{n+1}\xi(1-\xi)^{1+1/n}. \tag{2.24}$$

The correction satisfies $\mathcal{F}_Y(0) = 1$ and $\mathcal{F}_Y(1) = 0$, so that $q_{HB}(a)$ vanishes continuously as $a \to a_y^+$. In the limit $\tau_y \to 0$, the Herschel–Bulkley flux reduces to the power-law flux with the same $m$ and $n$.

## 3. Parallel-channel heat-transport model

### 3.1. Single-channel advection–conduction solution

For each hydraulic channel defined in Section 2, one-dimensional heat transport is considered along the semi-infinite channel $x_1 \ge 0$ for $t > 0$, with the outlet response monitored at the cross-section $x_1 = L$. The hydraulic quantities $q(a)$ and $\bar{u}(a)$ are provided by the closure for aperture $a$. Longitudinal conduction and in-channel Taylor dispersion are neglected, consistent with the classical single-fracture closure of Lauwerier (1955), while conductive exchange through both walls is represented by a semi-infinite matrix memory (Fig. 1b). The applicability of these assumptions to the spreading diagnostics used below is assessed after Eq. (3.6).

Let $\theta = (T_f - T_0)/(T_{\text{in}} - T_0)$ be the normalized fracture-fluid temperature, where $T_f$ is the fracture-fluid temperature, $T_0$ the initial uniform temperature of fluid and matrix, and $T_{\text{in}}$ the imposed inlet temperature, so that $\theta(x_1, 0 \mid a) = 0$ and $\theta(0, t \mid a) = 1$. Eliminating the wall-normal matrix-conduction problem gives, with $s$ the Laplace variable and hats denoting Laplace transforms in time, the Laplace-space channel balance

$$\bar{u}(a)\frac{\partial\hat{\theta}}{\partial x_1} + \left[s + \kappa(a)\sqrt{s}\right]\hat{\theta} = 0, \qquad \hat{\theta}(0, s \mid a) = \frac{1}{s}, \tag{3.1}$$

where

$$\kappa(a) = \frac{2\phi_m\sqrt{D_m}}{a}. \tag{3.2}$$

Here $D_m$ is the saturated-matrix thermal diffusivity and $\phi_m$ is the matrix-to-fracture-fluid volumetric heat-capacity ratio. Their definitions, the interface conditions, and the elimination leading to Eq. (3.1) are collected in Appendix D.

Solving Eq. (3.1) gives

$$\hat{\theta}(x_1, s \mid a) = \frac{1}{s}\exp\left[-\frac{x_1}{\bar{u}(a)}\left(s + \kappa(a)\sqrt{s}\right)\right]. \tag{3.3}$$

At $x_1 = L$, the response is

$$\hat{\theta}(s \mid a) = \frac{1}{s}\exp\left[-s\,t_a(a) - \lambda(a)\sqrt{s}\right], \tag{3.4}$$

where

$$t_a(a) = \frac{L}{\bar{u}(a)}, \qquad \lambda(a) = \kappa(a)t_a(a). \tag{3.5}$$

Equivalently, $t_a = La/q(a)$ and $\lambda = 2\phi_m\sqrt{D_m}L/q(a)$; larger-flux channels therefore have shorter transit times and weaker matrix memory.

Inverting the Laplace transform gives the single-channel breakthrough curve

$$\theta(t \mid a) = \operatorname{erfc}\left[\frac{\lambda(a)}{2\sqrt{t - t_a(a)}}\right] H\left(t - t_a(a)\right), \tag{3.6}$$

where $H$ is the Heaviside function. This response is the single-channel advection–conduction solution used in the aperture-distributed model below.

The lubrication closure retains the wall-normal velocity profile in the aperture-to-flux mapping $q(a)$ and advects heat at the aperture-averaged velocity $\bar{u}(a)$, reducing the response to the advection–matrix-exchange kernel studied here.

Over a channel transit, the longitudinal molecular-conduction length relative to $L$ is $\sqrt{2/Pe_L}$, where $Pe_L = \bar{u}(a)L/\alpha_f$ and $\alpha_f \approx 1.4\times10^{-7}\ \text{m}^2\,\text{s}^{-1}$ for aqueous fluids. At the flux-weighted mean velocity this ratio is at most of order $10^{-3}$. The estimate is not uniform near the Herschel–Bulkley yield threshold, but those classes carry negligible advective flux. Within Eq. (3.24), an omitted in-channel dispersion term would enter the channel-scale variance rather than the inter-channel selection signal, and it does not alter the late-time matrix-memory prefactors. It can, however, be comparable to the inter-channel contribution before wall-normal conduction homogenizes the aperture; at late times it accumulates with the mobile time, $O(t^{1/2})$, rather than with the $O(t)$ matrix-memory variance. With $t_c = L/\langle\bar{u}(a)\rangle_w$, this limit is

$$t_\perp = \frac{\langle a\rangle_p^2}{\alpha_f} \simeq 6\times10^{-2}\,t_c \tag{3.7}$$

for the reference parameters of Table 1. The spreading diagnostics are therefore interpreted only for $t \gtrsim t_\perp$; the two-dimensional solver of Appendix C uses the same depth-averaged closure.

Neglecting lateral heat exchange restricts the fracture the ensemble represents, not the ensemble itself, whose classes are aperture values rather than objects placed side by side. The assumption adopted is that matrix conduction moves no appreciable heat across the transverse scale on which the aperture, and hence the local velocity, changes, so each class exchanges heat with the matrix as if thermally isolated from the others. In a connected rough fracture that scale is the aperture correlation length, and for the fields of Section 7.6 the assumption holds over some tens of advective time scales. Beyond it, the model is an independent-channel end-member rather than a laterally coupled prediction. Because lateral conduction transfers heat from faster to slower paths, the model is expected to overestimate the inter-channel variance $\text{Var}_w[\mathcal{M}(t \mid a)]$ of the mean front position defined in Section 3.4. Quantifying that reduction would require path adjacency, spatial correlation, and an in-plane-conduction solver, none of which enters the present closure.

### 3.2. Flux-weighted parallel-channel ensemble

Consider hydraulically independent channels (Fig. 1a), each governed by the response of Section 3.1. The semi-infinite domain is monitored at $x_1 = L$; because advection is downstream and matrix exchange is local in $x_1$, its upstream solution coincides with that of a finite channel with an outflow boundary at $L$.

The channel apertures are described by a probability density $p(a)$ over geometric aperture classes. Fluid is injected continuously at the prescribed temperature $T_{\text{in}}$. All channels therefore have the same inlet temperature but carry aperture-dependent volumetric fluxes $q(a)$. The thermal power injected into channels with apertures in $[a, a + \mathrm{d}a]$ is proportional to $q(a)p(a)\,\mathrm{d}a$. Consequently, the macroscopic response is obtained by flux-weighted superposition of the single-channel solutions, with aperture classes weighted by $q(a)p(a)\,\mathrm{d}a$ rather than by their geometric frequency $p(a)\,\mathrm{d}a$.

Only hydraulically active channels contribute to the injected heat flux, so the superposition is restricted to the active aperture set $\mathcal{A}$. For Newtonian, power-law, and Ellis fluids, the full aperture support is active, $\mathcal{A} = (0, \infty)$. For Herschel–Bulkley fluids, the yield condition restricts the active set to $\mathcal{A} = (a_y, \infty)$. The corresponding active probability mass is

$$P_{\mathrm{act}} = \int_{\mathcal{A}} p(a)\,\mathrm{d}a. \tag{3.8}$$

Active-set integrals are written as

$$\langle f(a)\rangle_{\mathcal{A}} = \int_{\mathcal{A}} f(a)p(a)\,\mathrm{d}a. \tag{3.9}$$

They are not normalized by $P_{\mathrm{act}}$; this convention is used so that $\langle q(a)\rangle_{\mathcal{A}}$ is the flux per channel averaged over the whole aperture population, with inactive classes contributing zero.

The flux-weighted aperture density is

$$w(a) = \frac{q(a)p(a)}{\langle q(a)\rangle_{\mathcal{A}}}, \qquad a \in \mathcal{A}, \tag{3.10}$$

and satisfies $\int_{\mathcal{A}} w(a)\,\mathrm{d}a = 1$. The corresponding flux-weighted average is

$$\langle f(a)\rangle_w = \int_{\mathcal{A}} f(a)w(a)\,\mathrm{d}a = \frac{\langle q(a)f(a)\rangle_{\mathcal{A}}}{\langle q(a)\rangle_{\mathcal{A}}}. \tag{3.11}$$

The outlet response is

$$\theta_{\mathrm{out}}(t) = \langle \theta(t \mid a)\rangle_w, \qquad \hat{\theta}_{\mathrm{out}}(s) = \left\langle \hat{\theta}(s \mid a)\right\rangle_w. \tag{3.12}$$

Using the single-channel solution, Eq. (3.6), the outlet response in time is

$$\theta_{\mathrm{out}}(t) = \left\langle \mathrm{erfc}\left[\frac{\lambda(a)}{2\sqrt{t - t_a(a)}}\right] H\left(t - t_a(a)\right)\right\rangle_w. \tag{3.13}$$

The corresponding Laplace-space response is

$$\hat{\theta}_{\mathrm{out}}(s) = \frac{1}{s}\left\langle \exp\left[-s\,t_a(a) - \lambda(a)\sqrt{s}\right]\right\rangle_w. \tag{3.14}$$

Since all channels share the inlet temperature, injected thermal power and volumetric flux are proportional, so $\theta_{\mathrm{out}}$ is the flow-averaged, or mixing-cup, outlet temperature: the quantity a flow-weighted measurement returns, and the quantity the two-dimensional reference solver of Appendix C computes. It differs from a cross-section-averaged temperature whenever part of the aperture population is inactive. Rheology enters through both the single-channel transport parameters $t_a(a)$ and $\lambda(a)$, and the flux-weighted sampling measure $w(a)$.

At late times, expansion of the complementary error function in Eq. (3.6) gives the single-channel tail

$$1 - \theta(t \mid a) \sim \frac{\lambda(a)}{\sqrt{\pi}}\,t^{-1/2}, \qquad t \to \infty, \tag{3.15}$$

so that $\lambda(a)$ sets the amplitude of the matrix-controlled $t^{-1/2}$ decay of each channel. Because the superposition is linear, flux-weighted averaging of Eq. (3.15) gives the ensemble tail, whose amplitude defines the late-time breakthrough prefactor

$$1 - \theta_{\mathrm{out}}(t) \sim P_{\mathrm{BTC}}^{(\infty)}\,t^{-1/2}, \qquad P_{\mathrm{BTC}}^{(\infty)} = \frac{\langle \lambda(a)\rangle_w}{\sqrt{\pi}} = \frac{2\phi_m\sqrt{D_m}\,L\,P_{\mathrm{act}}}{\sqrt{\pi}\,\langle q(a)\rangle_{\mathcal{A}}}, \tag{3.16}$$

where the second equality follows from the definition of $\lambda(a)$ and Eq. (3.11).

### 3.3. Classical planar-channel reference for non-Newtonian transport

The single-channel solution, Eq. (3.6), coincides with the classical planar-fracture solution of Lauwerier (1955) under the identification of $t_a(a)$ and $\lambda(a)$ established in Appendix D, with the rheology entering only through $q(a)$. It therefore provides a classical planar-channel reference that is valid for each constitutive law considered here, without requiring a Newtonian limit. For each rheology it is evaluated at the mean aperture $a_0 = \langle a\rangle_p$, the mean over the aperture density $p(a)$, and at the same flux-weighted mean velocity $\langle \bar{u}(a)\rangle_w$ as the heterogeneous ensemble. Only two additional diagnostics are introduced:

$$E_{\mathrm{BTC}} = \max_{10^{-5} \le \theta_{\mathrm{L}} \le 1-10^{-5}} \left|\theta_{\mathrm{out}} - \theta_{\mathrm{L}}\right|, \qquad R_\infty = \frac{P_{\mathrm{BTC}}^{(\infty)}}{P_{\mathrm{BTC,L}}^{(\infty)}} = \frac{a_0\langle \bar{u}(a)\rangle_w P_{\mathrm{act}}}{\langle q(a)\rangle_{\mathcal{A}}}, \tag{3.17}$$

where $\theta_{\mathrm{L}}$ and $P_{\mathrm{BTC,L}}^{(\infty)}$ denote, respectively, the response and late-time prefactor of the homogeneous single-channel reference. The first diagnostic tests the complete breakthrough transition, whereas the second isolates its late-time amplitude.

Fig. 2 maps these diagnostics against $\mathrm{CV}_a$ and the imposed gradient $G = |\partial_{x_1}\mathcal{P}|$, using $\mathcal{G}_E = Ga_0/(2\tau_{1/2})$ and $\mathcal{G}_{HB} = Ga_0/(2\tau_y) = 1/\beta_y$, with $a_0 = \langle a\rangle_p$ and $\beta_y = a_y/\langle a\rangle_p$ the yield aperture of Eq. (2.20) normalized by the mean aperture. The Ellis panels use fluid F2 of Table 1; the Herschel–Bulkley panels use $n = 0.61$, $\tau_y = 0.20$ Pa, and $m = 9.23$ Pa s$^n$. Here the fluid properties are fixed and $G$ is swept, whereas the deterministic cases of Section 6.1 prescribe $\beta_y$ and adjust the dimensional loading to hold the reference thermal Péclet number $Pe_{\mathrm{th}} = \rho_f c_{p,f}\langle \bar{u}(a)\rangle_w\langle a\rangle_p/k_r$ of Eq. (6.1) fixed.

For $\mathrm{CV}_a = 0$ the flux-weighted density degenerates to a single aperture class and the ensemble reduces analytically to the classical solution. The computed values, $E_{\mathrm{BTC}} < 5.0 \times 10^{-16}$ and $R_\infty = 1$ to numerical precision, hold for every rheology, including those whose aperture-to-flux mapping is not a monomial. Since the limit is exact by construction, this verifies the quadrature and the normalization rather than the superposition, which is instead tested against resolved two-dimensional simulations in Section 7.6. With increasing $\mathrm{CV}_a$, Ellis departures are amplified once the shear-thinning branch dominates, whereas the Herschel–Bulkley response changes most strongly near partial yielding. The interior of the maps is therefore a result: it quantifies how far heterogeneous non-Newtonian flux partitioning displaces the response from the classical prediction, without imposing a universal adequacy tolerance.

The maps retain the untruncated lognormal support. Their red contour marks 99% retention of the third flux moment below the reference cutoff; its application-dependent interpretation and the separate Sobol-domain restriction are discussed in Section 6.2.

### 3.4. Longitudinal moments of the thermal front

In addition to the outlet breakthrough curve, the spatial propagation of the thermal disturbance is characterized through the longitudinal front density

$$p_x(x_1, t \mid a) = -\frac{\partial\theta(x_1, t \mid a)}{\partial x_1}. \tag{3.18}$$

For the advection–conduction solution, $p_x$ is used as a normalized diagnostic density for the position of the thermal front along the channel, since $\theta(0, t \mid a) = 1$ and $\theta(x_1, t \mid a) \to 0$ as $x_1 \to \infty$. The ensemble density is formed with the same flux-weighted average as the outlet response, Eq. (3.11), so that the moments below describe the front carried by the injected thermal signal rather than the thermal energy stored per unit fracture area; a volume-weighted average would give different amplitudes for the same channel solutions. With that convention,

$$p_x(x_1, t) = \langle p_x(x_1, t \mid a)\rangle_w. \tag{3.19}$$

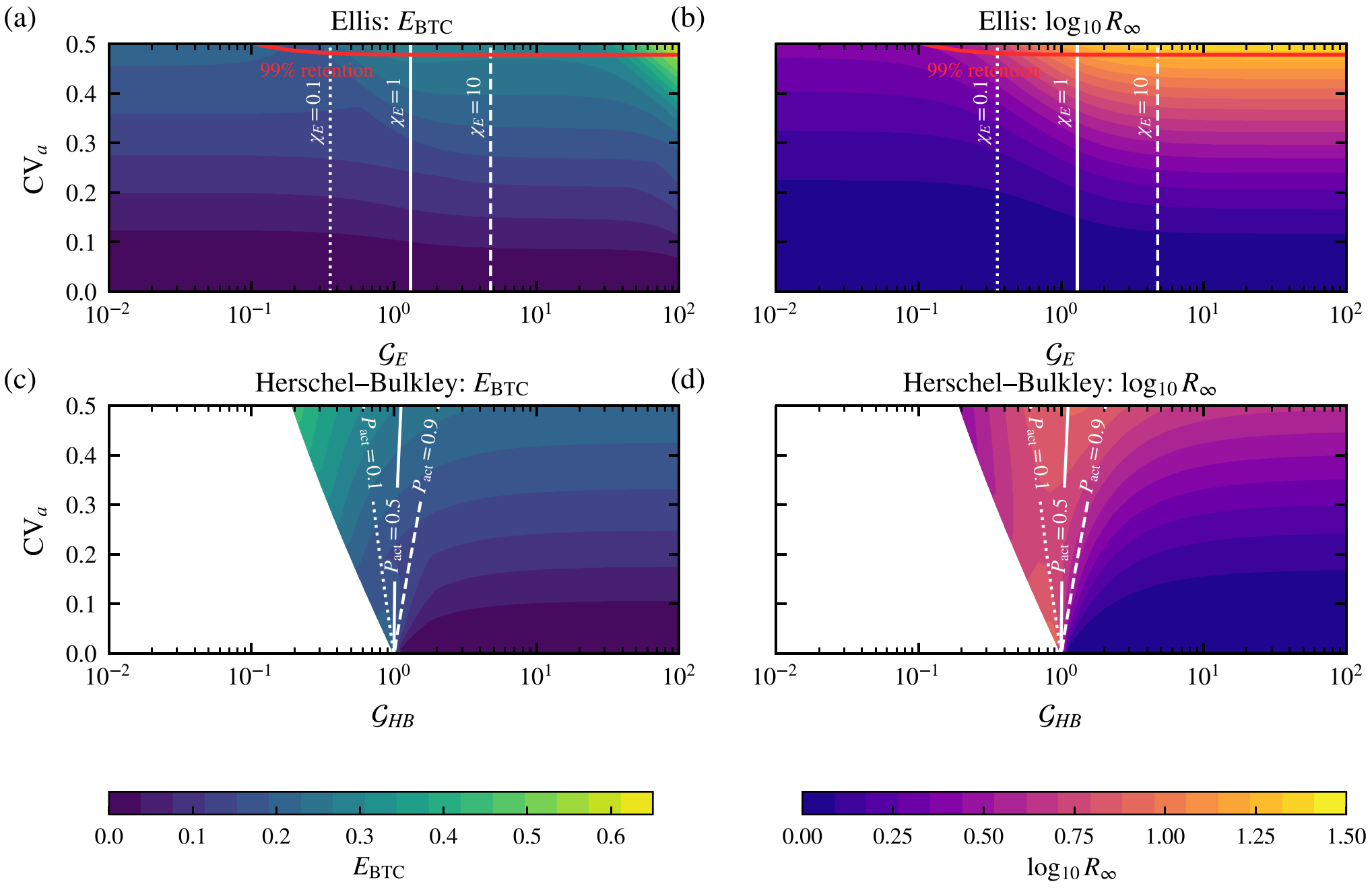

**Fig. 2.** Comparison with the classical planar-channel Lauwerier solution at the same mean velocity. The top row uses Ellis fluid F2 and the bottom row the Herschel–Bulkley rheology specified in the text. Panels (a,c) show $E_{\mathrm{BTC}}$ and panels (b,d) show $\log_{10} R_\infty$, for which zero denotes equal late-time amplitudes. The abscissae are $\mathcal{G}_E = Ga_0/(2\tau_{1/2})$ and $\mathcal{G}_{HB} = Ga_0/(2\tau_y)$, with $a_0 = \langle a \rangle_p$. White contours mark $\chi_E = 0.1, 1, 10$ in the Ellis panels and $P_{\mathrm{act}} = 0.1, 0.5, 0.9$ in the Herschel–Bulkley panels; white regions with $P_{\mathrm{act}} < 10^{-4}$ are excluded. The red contour marks 99% retention of the third flux moment below $a_{\max} = 5.0 \times 10^{-2}$ m, on the side containing $\mathrm{CV}_a = 0$. The classical solution is recovered to machine precision along $\mathrm{CV}_a = 0$ whenever the channel is yielded. The Sobol domain is delimited separately in Section 6.2.

The first two longitudinal moments are defined by

$$\mathcal{M}_\ell(t \mid a) = \int_0^\infty x_1^\ell \, p_x(x_1, t \mid a)\, \mathrm{d}x_1, \qquad \ell = 1, 2, \tag{3.20}$$

and, by linearity,

$$\mathcal{M}_\ell(t) = \langle \mathcal{M}_\ell(t \mid a) \rangle_w, \qquad \ell = 1, 2. \tag{3.21}$$

Throughout, the first moment is written without its index, $\mathcal{M} \equiv \mathcal{M}_1$, and the longitudinal variance of the front is denoted by $\mathcal{V}$.

The Laplace-space moments and their exact inverse transforms are collected in Appendix D. Those closed forms are used for all finite-time calculations; the main text retains only their variance decomposition and the limiting expressions needed for interpretation.

The longitudinal variance is

$$\mathcal{V}(t) = \mathcal{M}_2(t) - \mathcal{M}(t)^2. \tag{3.22}$$

For a fixed aperture, define the channel-wise variance as

$$\sigma_x^2(a, t) = \mathcal{M}_2(t \mid a) - \mathcal{M}(t \mid a)^2. \tag{3.23}$$

The law of total variance gives

$$\mathcal{V}(t) = \left\langle \sigma_x^2(a, t) \right\rangle_w + \mathrm{Var}_w[\mathcal{M}(t \mid a)]. \tag{3.24}$$

Here $\mathrm{Var}_w$ denotes the variance with respect to the flux-weighted density $w(a)$; the covariance $\mathrm{Cov}_w$ used in Appendix D is defined analogously. The first term is the flux-weighted channel-scale spreading generated by fracture–matrix exchange. The second term is the persistent inter-channel contribution caused by aperture-dependent differences in mean front position.

The moments in Eqs. (3.20) and (3.21) are defined on the semi-infinite reference domain and quantify front propagation independently of the monitoring location $L$. They should therefore not be interpreted as moments of the thermal disturbance retained within the finite interval $0 \le x_1 \le L$. Because the semi-infinite and finite-channel solutions coincide pointwise upstream of the monitoring section, these moments provide natural pre-outlet diagnostics. Comparisons with finite-domain simulations are accordingly restricted to regimes in which breakthrough at $x_1 = L$ is negligible. After significant breakthrough, $\theta_{\mathrm{out}}(t)$ is the relevant finite-channel observable, whereas the semi-infinite moments continue to characterize unconstrained front propagation; consequently, $\mathcal{M}(t)/L$ is not bounded by unity.

*Short-time limit.* For times short compared with the matrix-exchange time of the active channels, the leading ensemble behavior is

$$\mathcal{M}(t) \sim \langle \bar{u}(a) \rangle_w\, t, \qquad \mathcal{V}(t) \sim \mathrm{Var}_w[\bar{u}(a)] t^2. \tag{3.25}$$

Thus, the leading short-time variance is ballistic and reflects persistent velocity contrasts between aperture classes. The channel expansions and the $t^{5/2}$ matrix-exchange correction are given in Appendix D.

*Late-time limit.* At late times, the semi-infinite matrix memory fixes the temporal scaling. From the exact inversions in Eqs. (D.6) and (D.7), each channel satisfies

$$\mathcal{M}(t \mid a) \sim \frac{2}{\sqrt{\pi}} \frac{\bar{u}(a)}{\kappa(a)} t^{1/2}, \qquad \sigma_x^2(a, t) \sim 2\left(1 - \frac{2}{\pi}\right)\left[\frac{\bar{u}(a)}{\kappa(a)}\right]^2 t. \tag{3.26}$$

Using the definitions of $\bar{u}(a)$ and $\kappa(a)$ in Eqs. (2.5) and (3.2), define $X(a) = \bar{u}(a)/\kappa(a) = q(a)/(2\phi_m\sqrt{D_m})$. Flux-weighted averaging of the first limit in Eq. (3.26) gives the ensemble front advance $\mathcal{M}(t) \sim P_{\mathcal{M}}^{(\infty)} t^{1/2}$, with $P_{\mathcal{M}}^{(\infty)} = (2/\sqrt{\pi}) \langle X(a) \rangle_w$. The late-time variance becomes

$$\mathcal{V}(t) \sim P_{\mathcal{V}}^{(\infty)} t, \qquad P_{\mathcal{V}}^{(\infty)} = P_{\mathrm{mem}}^{(\infty)} + P_{\mathrm{inter}}^{(\infty)}, \tag{3.27}$$

with

$$P_{\mathrm{mem}}^{(\infty)} = 2\left(1 - \frac{2}{\pi}\right)\left\langle X(a)^2 \right\rangle_w, \qquad P_{\mathrm{inter}}^{(\infty)} = \frac{4}{\pi} \mathrm{Var}_w[X(a)]. \tag{3.28}$$

The first contribution is the intrinsic matrix-memory spreading within individual channels, whereas the second is the persistent inter-channel contribution.

Using the active-set integrals defined in Eq. (3.9), the total late-time prefactor can be written directly in terms of the aperture-dependent

flux as

$$P_{\mathcal{V}}^{(\infty)} = \frac{1}{2\phi_m^2 D_m}\left[\frac{\langle q(a)^3\rangle_{\mathcal{A}}}{\langle q(a)\rangle_{\mathcal{A}}} - \frac{2}{\pi}\left(\frac{\langle q(a)^2\rangle_{\mathcal{A}}}{\langle q(a)\rangle_{\mathcal{A}}}\right)^2\right]. \tag{3.29}$$

## 4. Analytical limits for rheological channel selection

This section gives the semi-analytical late-time amplitudes obtained by substituting each rheology-dependent flux law into the active-set flux moments derived above. The aim is to expose which aperture moments control outlet tailing, front advance, and longitudinal spreading for each constitutive closure.

When closed-form aperture moments are required, the aperture density $p(a)$ is assumed to be lognormal. This distribution is commonly used in stochastic descriptions of single-fracture aperture variability and is consistent with reported aperture statistics for rough fractures (Moreno et al., 1988; Tsang et al., 1988; Hakami and Larsson, 1996). It has positive support, can be parameterized directly in terms of the arithmetic mean aperture and coefficient of variation, and admits closed-form raw moments. Its probability density is

$$p(a) = \frac{1}{a\sigma\sqrt{2\pi}}\exp\left[-\frac{(\ln a-\mu)^2}{2\sigma^2}\right], \qquad a > 0, \tag{4.1}$$

where $\mu$ and $\sigma$ are the mean and standard deviation of $\ln a$, respectively.

The average of a function $f(a)$ with respect to the full aperture distribution is $\langle f(a)\rangle_p = \int_0^\infty f(a)\,p(a)\,\mathrm{d}a$. The corresponding lognormal raw moments are

$$\langle a^k\rangle_p = \exp\left(k\mu + \frac{1}{2}k^2\sigma^2\right), \qquad k \in \mathbb{R}. \tag{4.2}$$

For the Newtonian, power-law, and Ellis closures, the active aperture set coincides with the full support, $\mathcal{A} = (0,\infty)$, so that $\langle a^k\rangle_{\mathcal{A}} = \langle a^k\rangle_p$. For Herschel–Bulkley fluids, the same aperture density is used, but the active-set integrals are truncated over $\mathcal{A} = (a_y,\infty)$.

The arithmetic mean aperture and coefficient of variation are

$$\langle a\rangle_p = \exp\left(\mu + \frac{\sigma^2}{2}\right), \qquad \mathrm{CV}_a = \sqrt{\exp(\sigma^2)-1}. \tag{4.3}$$

Equivalently, the lognormal parameters are $\sigma = \sqrt{\ln(1+\mathrm{CV}_a^2)}$ and $\mu = \ln\langle a\rangle_p - \sigma^2/2$.

### 4.1. Newtonian reference

For $q_N = C_N a^3$, Eqs. (3.16), (3.26), and (3.29) show that outlet tailing, front advance, and longitudinal variance depend, respectively, on the raw aperture moments of orders 3, (3, 6), and (3, 6, 9). The corresponding closed-form lognormal prefactors are listed in Appendix D. This hierarchy is retained here because it identifies the progressively stronger upper-tail weighting of the three transport diagnostics.

### 4.2. Power-law shear-thinning limit

For the power-law closure, the aperture exponent is defined as

$$\beta_{PL} = 2 + \frac{1}{n}, \tag{4.4}$$

so that $q_{PL} = C_{PL}a^{\beta_{PL}}$. The same three diagnostics therefore sample moments of orders $\beta_{PL}$, $2\beta_{PL}$, and $3\beta_{PL}$. Since $\beta_{PL} > 3$ for shear-thinning flow, each diagnostic weights the large-aperture tail more strongly than its Newtonian counterpart. The longitudinal variance, which samples the highest of these orders, is correspondingly the most sensitive to the upper end of the aperture distribution, and the finite-support considerations of Section 3.3 apply to it most strongly. The explicit prefactors and their lognormal forms are given in Appendix D; Section 5 retains the reduced combinations needed to interpret the results.

### 4.3. Ellis transition

The Ellis closure combines the Newtonian and shear-thinning branches, so the three master prefactors, $P_{\mathrm{BTC}}^{(\infty)}$, $P_{\mathcal{M}}^{(\infty)}$, and $P_{\mathcal{V}}^{(\infty)}$, depend on mixed flux moments. For $r = 1, 2, 3$ these are

$$\langle q_E(a)^r\rangle_{\mathcal{A}} = \sum_{j=0}^{r}\binom{r}{j}C_N^{r-j}C_{PL}^{j}\left\langle a^{3(r-j)+j\beta_{PL}}\right\rangle_{\mathcal{A}}, \qquad r = 1, 2, 3. \tag{4.5}$$

The mixed terms in Eq. (4.5) show that the Ellis late-time amplitudes are not simple linear interpolations between the Newtonian and power-law amplitudes. The explicit lognormal expansion and its substitution into the master prefactors are given in Appendix D.

### 4.4. Herschel–Bulkley yield-controlled selection

For the Herschel–Bulkley closure, the late-time amplitudes are controlled by yielded flux moments over $\mathcal{A} = (a_y,\infty)$: in particular, $P_{\mathrm{BTC},HB}^{(\infty)}$ follows from Eq. (3.16) with $q = q_{HB}$, and the factor $P_{\mathrm{act}} = \langle 1\rangle_{\mathcal{A}}$ is retained because active-set integrals are not renormalized. The mean and variance follow directly from the same second and third flux moments used in Eqs. (3.26) and (3.29). Because $q_{HB}$ vanishes continuously at $a_y$, $\lambda \propto q_{HB}^{-1}$ diverges at the yield edge while the aperture density remains finite there; near-yield channels therefore introduce a slowly decaying relative correction of order $t^{-n/[2(n+1)]}$ to the leading outlet tail. This asymptotic signature can lie beyond operational time scales and is used here to characterize the mathematical approach to the matrix-controlled limit, not as a guaranteed field-time observable. The yielded lognormal quadrature and the $a_y \to 0$ recovery of the power-law moments are given in Appendix D.

## 5. Similarity structure of the late-time amplitudes

The closed-form limits above reveal three related similarity structures. The monomial Newtonian and power-law closures admit a one-parameter collapse of their reduced amplitudes, the Ellis closure introduces a moment-dependent two-branch crossover, and the Herschel–Bulkley closure adds the standardized position of the yield edge.

For the full-support monomial closures,

$$q(a) = Ca^{\beta}, \tag{5.1}$$

the late-time prefactors in Eqs. (3.16), (3.26), and (3.29) reduce to lognormal moment ratios. Using Eq. (4.2), these lognormal factors collapse onto the single similarity variable

$$g = \beta^2\sigma^2. \tag{5.2}$$

Here $\beta = 3$ and $C = C_N$ for the Newtonian cubic law, whereas $\beta = \beta_{PL} = 2 + 1/n$ and $C = C_{PL}$ for the power-law closure. Since $\exp(\mu)$ is the median aperture of the lognormal distribution, the corresponding median-aperture hydraulic scale is

$$q_\beta = C\exp(\beta\mu). \tag{5.3}$$

Factoring out $q_\beta$ and its corresponding powers from the dimensional prefactors gives

$$\widetilde{P}_{\mathrm{BTC}}^{(\infty)} \equiv \frac{\sqrt{\pi}\,q_\beta}{2\phi_m\sqrt{D_m}\,L}P_{\mathrm{BTC}}^{(\infty)} = \exp\left(-\frac{g}{2}\right), \tag{5.4}$$

$$\widetilde{P}_{\mathcal{M}}^{(\infty)} \equiv \frac{\phi_m\sqrt{\pi D_m}}{q_\beta}P_{\mathcal{M}}^{(\infty)} = \exp\left(\frac{3g}{2}\right), \tag{5.5}$$

$$\widetilde{P}_{\mathcal{V}}^{(\infty)} \equiv \frac{2\phi_m^2 D_m}{q_\beta^2}P_{\mathcal{V}}^{(\infty)} = \exp(3g)\left[\exp(g) - \frac{2}{\pi}\right]. \tag{5.6}$$

Relative to the Newtonian exponent $\beta = 3$, the power-law closure is equivalent, at the level of reduced amplitudes, to a Newtonian response with effective log-aperture width

$$\sigma_{\mathrm{eff}} = \frac{\beta_{PL}}{3}\sigma = \frac{2n+1}{3n}\sigma, \tag{5.7}$$

so that shear thinning amplifies the apparent aperture variability. The normalization-dependent hydraulic scale $q_\beta$ cancels in the late-time relative front width,

$$\lim_{t\to\infty} \frac{\mathcal{V}(t)}{\mathcal{M}(t)^2} = \frac{\pi}{2}\exp(g) - 1, \tag{5.8}$$

which is therefore an exact function of $g$ alone.

The Ellis and Herschel–Bulkley closures extend this collapse by one coordinate each; their reduced forms are collected in Appendix D. The Ellis flux carries two median-aperture hydraulic scales, $C_N \exp(3\mu)$ and $C_{PL}\exp(\beta_{PL}\mu)$, and the added coordinate is their ratio $\tilde{\chi}_E = \chi_E \exp[-(\beta_{PL} - 3)\sigma^2/2]$, with $\chi_E$ the arithmetic-mean branch ratio defined in Eq. (2.18). Together with $g_N = 9\sigma^2$ and $g_{PL} = \beta_{PL}^2\sigma^2$, the dimensionless function $m_r(\tilde{\chi}_E; g_N, g_{PL})$ fixes the reduced late-time prefactors. Higher-order flux moments become shear-thinning-dominated at smaller $\tilde{\chi}_E$, because they weight more strongly the large-aperture tail, where the shear-thinning branch grows faster than the Newtonian one. For Herschel–Bulkley flow the added coordinate is the standardized yield-edge position $\tilde{y} = (\ln a_y - \mu)/\sigma$, with active probability $P_{\text{act}} = 1-\Phi(\tilde{y})$ and $\Phi$ the standard normal cumulative distribution; the reduced response is controlled by $(\beta_{PL}, \sigma, \tilde{y})$ through a dimensionless yielded-moment quadrature $\mathcal{G}_r$, and the physical variables $(\text{CV}_a, n, \beta_y)$ span the same similarity space.

## 6. Numerical implementation and sensitivity metrics

### *6.1. Fluid, matrix, and aperture parameters*

Calculations were performed using the parameter set summarized in Table 1. Fluid and matrix thermal properties were kept fixed, while aperture heterogeneity and rheological parameters were varied according to the cases listed in the table. The Ellis parameters for silicone oil, xanthan gum, and 1000 wppm (parts per million by weight) HPAM are based on the rheological characterizations of Uddin et al. (2012), Lenci et al. (2022b), and Habibpour and Clark (2017), respectively. The silicone-oil data of Uddin et al. (2012) are reported there as a Carreau fit; the $(\eta_0, \tau_{1/2}, n)$ values used here retain its zero-shear viscosity and high-shear thinning branch, expressed in the Ellis form of Appendix A.

The advective scale was fixed by prescribing the reference thermal Péclet number

$$Pe_{\text{th}} = \frac{\rho_f c_{p,f} \langle \bar{u}(a)\rangle_w \langle a\rangle_p}{k_r}. \tag{6.1}$$

For each rheological case, the imposed pressure-gradient magnitude was adjusted to keep $Pe_{\text{th}}$ fixed.

The normalized diagnostics depend on the aperture statistics, the constitutive shape parameters, and the reference memory–advection group

$$\Lambda_0 = \frac{4\phi_m L}{Pe_{\text{th}}\langle a\rangle_p}, \tag{6.2}$$

Built on $Pe_{\text{th}}$ and the flux-weighted mean velocity, $\Lambda_0$ sets a reference scale rather than the local ratio of any aperture class. The latter is

$$\frac{\lambda(a)^2}{t_c} = \Lambda_0 \left(\frac{\langle a\rangle_p}{a}\right)^2 \left[\frac{\langle \bar{u}(a)\rangle_w}{\bar{u}(a)}\right]^2, \tag{6.3}$$

and equals $\Lambda_0$ at $a = \langle a\rangle_p$ only if that channel carries the flux-weighted mean velocity. The velocity factor spans more than three orders of magnitude over the deterministic set and diverges at the mean aperture for Herschel–Bulkley cases with $\beta_y \geq 1.25$, where that channel is unyielded. The common rescaling of Section 6.2 holds $Pe_{\text{th}}$ and hence $\Lambda_0$ fixed, so the normalized curves depend on constitutive shape rather than flux amplitude. Here $\Lambda_0 = 2.01 \times 10^2$; the consistency $m$ affects only the dimensional loading required to realize each case.

Table 2 reports the dimensional loadings removed by this normalization. Unlike Section 3.3, which fixes $\tau_y$ and sweeps $G$, these cases prescribe $\beta_y$ and derive $\tau_y$ from the gradient required by $Pe_{\text{th}}$. The mapping is given in Appendix A.1; Ellis F1 and $\beta_y = 2.00$ are limiting illustrations rather than operating conditions.

For $\beta_y = 2.00$, $P_{\text{act}} \approx 1.6 \times 10^{-4}$ at $\text{CV}_a = 0.20$ and $2.3 \times 10^{-2}$ at $\text{CV}_a = 0.40$, so a finite fracture would contain few mobilized paths. The case is retained only as a limiting illustration; the GSA is capped at $\log_{10}\beta_y = 0.15$ ($\beta_y \approx 1.4$).

### *6.2. Diagnostics and global sensitivity analysis*

A variance-based GSA screens the reduced model over the prescribed ranges of Table 3; its rankings are conditional on that domain. The physical inputs are $\text{CV}_a$ and $n$, together with the Ellis branch ratio $\chi_E$ or the Herschel–Bulkley yield position $\beta_y$. Hydraulic outputs use the unscaled flux laws, whereas transport outputs are normalized to the same $Pe_{\text{th}}$.

Following the full-support cutoff diagnostic of Section 3.3, the Sobol domain was restricted using the most tail-sensitive output, the third flux moment. Its retained fraction for a lognormal population is

$$R_{q^3} = \Phi\left[\frac{\ln(a_{\max}/\langle a\rangle_p) - \left(3(2+1/n) - \frac{1}{2}\right)\sigma^2}{\sigma}\right], \qquad \sigma^2 = \ln(1+\text{CV}_a^2). \tag{6.4}$$

For $n \geq 0.20$ and $\text{CV}_a \leq 0.38$, the minimum retained fraction is $R_{q^3} = 0.9991$ at $(n, \text{CV}_a) = (0.20, 0.38)$. Truncating and renormalizing the aperture density changes the prefactors by at most about $9\times10^{-4}$ at the least favorable corner; the GSA therefore retains the full lognormal support while reporting this tail sensitivity explicitly.

For the Ellis model, the input vector was

$$\boldsymbol{X}_E = \left(\text{CV}_a,\, n,\, \log_{10}\chi_E\right), \tag{6.5}$$

where $\chi_E$ is the Ellis branch ratio of Eq. (2.18), evaluated at the mean aperture, $a_0 = \langle a\rangle_p$, so that it measures the ratio between the shear-thinning and Newtonian flux contributions there. For the Herschel–Bulkley model, the input vector was

$$\boldsymbol{X}_{HB} = \left(\text{CV}_a,\, n,\, \log_{10}\beta_y\right). \tag{6.6}$$

For the transport calculations, each realization was normalized to the same reference thermal Péclet number. Let $q_0(a)$ denote the unscaled flux law and write

$$q(a) = S q_0(a). \tag{6.7}$$

Imposing the target flux-weighted mean velocity gives

$$S = u_{\text{target}} \frac{\langle q_0(a)\rangle_{\mathcal{A}}}{\left\langle q_0(a)^2/a\right\rangle_{\mathcal{A}}}, \tag{6.8}$$

where $u_{\text{target}} = Pe_{\text{th}} k_r/(\rho_f c_{p,f}\langle a\rangle_p)$ is the flux-weighted mean velocity associated with the prescribed thermal Péclet number. This normalization removes variations in the overall flux magnitude while preserving the rheology-induced redistribution of flux among aperture classes.

For monomial laws and Herschel–Bulkley flow at fixed $\beta_y$, this rescaling is equivalent to changing $G$ (and scaling $\tau_y$ consistently). In the Ellis law, however, $C_N \propto G$ and $C_{PL} \propto G^{1/n}$, so $\chi_E \propto \eta_0 m^{-1/n} G^{1/n-1}$: changing $G$ alone changes both branch balance and amplitude. The common factor $S$ instead leaves $\chi_E$ unchanged and is realizable at fixed $G$, for example through $\eta_0 \mapsto \eta_0/S$ and $m \mapsto mS^{-n}$. Thus the $\log_{10}\chi_E$ indices measure branch balance, not gradient magnitude.

The transport outputs were $\log P_{\text{BTC}}^{(\infty)}$, $\log P_{\mathcal{V}}^{(0)}$, $\log P_{\mathcal{M}}^{(\infty)}$, $\log P_{\mathcal{V}}^{(\infty)}$, $\log P_{\text{mem}}^{(\infty)}$, $\log P_{\text{inter}}^{(\infty)}$, and the fraction $F_{\text{inter}} = P_{\text{inter}}^{(\infty)}/P_{\mathcal{V}}^{(\infty)}$. Here $P_{\mathcal{V}}^{(0)} = \text{Var}_w[\bar{u}(a)]$ is the formal short-time coefficient of the reduced model, not a Taylor-dispersion prediction; it is subject to the $t \gtrsim t_\perp$ interpretation stated in Section 3.1. Moreover, $P_{\mathcal{M}}^{(0)} = \langle \bar{u}(a)\rangle_w$ is fixed and used only as a consistency check. Positive prefactors were log-transformed; the bounded $F_{\text{inter}}$ was analyzed linearly.

**Table 1**
Geometric, thermal, aperture-distribution, and rheological parameters used in the deterministic calculations.

| Set | Quantity | Symbol | Value(s) |
|---|---|---|---|
| Medium | Channel length | $L$ | 10 m |
| | Mean aperture | $\langle a\rangle_p$ | $1.0\times10^{-3}$ m |
| | Aperture distribution | $p(a)$ | Lognormal |
| | Aperture coefficient of variation | $\mathrm{CV}_a$ | 0.20, 0.40 |
| Thermal | Saturated matrix-to-fluid heat-capacity ratio | $\phi_m$ | 0.503 |
| | Effective saturated-matrix thermal diffusivity | $D_m$ | $1.66\times10^{-6}$ m$^2$ s$^{-1}$ |
| | Effective matrix thermal conductivity | $k_r$ | 3.5 W m$^{-1}$ K$^{-1}$ |
| | Fluid density | $\rho_f$ | 1000 kg m$^{-3}$ |
| | Fluid specific heat capacity | $c_{p,f}$ | 4189 J kg$^{-1}$ K$^{-1}$ |
| | Matrix porosity | $\phi_r$ | 0.10 |
| | Solid-rock density | $\rho_r$ | 2500 kg m$^{-3}$ |
| | Solid-rock specific heat capacity | $c_{p,r}$ | 750 J kg$^{-1}$ K$^{-1}$ |
| | Thermal Péclet number | $Pe_{\mathrm{th}}$ | 100 |
| Newtonian | Dynamic viscosity | $\eta$ | $1.0\times10^{-3}$ Pa s |
| Power-law | Consistency index | $m$ | 0.16 Pa s$^n$ |
| | Flow index | $n$ | 0.61 |
| Ellis | Newtonian limit (water) | $(\eta_0, \chi_E)$ | $\eta_0 = 1.0\times10^{-3}$ Pa s; $\chi_E = 0$ (shear-thinning branch suppressed) |
| | Silicone oil (F1) | $(\eta_0, \tau_{1/2}, n)$ | $\eta_0 = 9.75$ Pa s; $\tau_{1/2} = 8.48$ Pa; $n = 0.61$ |
| | Xanthan gum (F2) | $(\eta_0, \tau_{1/2}, n)$ | $\eta_0 = 4.42$ Pa s; $\tau_{1/2} = 0.20$ Pa; $n = 0.36$ |
| | 1000 wppm HPAM (F3) | $(\eta_0, \tau_{1/2}, n)$ | $\eta_0 = 3.51$ Pa s; $\tau_{1/2} = 0.35$ Pa; $n = 0.21$ |
| HB | Consistency index | $m$ | 0.16 Pa s$^n$ |
| | Flow index | $n$ | 0.61 |
| | Normalized yield aperture | $\beta_y$ | 0, 0.50, 1.25, 2.00 |

**Table 2**
Dimensional loading implied by each deterministic case at the prescribed $Pe_{\mathrm{th}} = 100$, $\langle a\rangle_p = 1.0\times10^{-3}$ m and $L = 10$ m, for which $\langle \bar{u}(a)\rangle_w = 8.36\times10^{-2}$ m s$^{-1}$. Entries are given as $\mathrm{CV}_a = 0.20/0.40$. Here $G$ is the imposed pressure-gradient magnitude, $\Delta\mathcal{P} = GL$ the pressure drop, $\tau_w = G\langle a\rangle_p/2$ the mean-aperture wall stress, and $\tau_y = G\beta_y\langle a\rangle_p/2$ the implied yield stress.

| Case | $G$ [Pa m$^{-1}$] | $\Delta\mathcal{P}$ [MPa] | $\tau_w$ [Pa] | $\tau_y$ [Pa] |
|---|---|---|---|---|
| Newtonian (water) | $7.6\times10^2/3.5\times10^2$ | 0.0076/0.0035 | 0.38/0.18 | – |
| Ellis F1 (silicone oil) | $6.6\times10^5/3.0\times10^5$ | 6.58/3.02 | 329/151 | – |
| Ellis F2 (xanthan gum) | $9.7\times10^3/3.9\times10^3$ | 0.097/0.039 | 4.9/2.0 | – |
| Ellis F3 (1000 wppm HPAM) | $3.2\times10^3/9.7\times10^2$ | 0.032/0.010 | 1.6/0.49 | – |
| Power-law ($\beta_y = 0$) | $1.2\times10^4/5.5\times10^3$ | 0.121/0.055 | 6.0/2.8 | 0/0 |
| HB $\beta_y = 0.50$ | $2.1\times10^4/6.2\times10^3$ | 0.213/0.062 | 10.7/3.1 | 5.3/1.5 |
| HB $\beta_y = 1.25$ | $4.2\times10^4/6.2\times10^3$ | 0.416/0.062 | 20.8/3.1 | 26.0/3.9 |
| HB $\beta_y = 2.00$ | $4.8\times10^4/5.6\times10^3$ | 0.480/0.056 | 24.0/2.8 | 48.0/5.6 |

**Table 3**
Input ranges used in the Sobol GSA. The parameters $\chi_E$ and $\beta_y$ were sampled uniformly in logarithmic space.

| Model | Parameter | Symbol | Range |
|---|---|---|---|
| Ellis | Aperture coefficient of variation | $\mathrm{CV}_a$ | [0.10, 0.38] |
| | Flow index | $n$ | [0.20, 1.00] |
| | Shear-thinning intensity | $\log_{10}\chi_E$ | [−2, 2] |
| Herschel–Bulkley | Aperture coefficient of variation | $\mathrm{CV}_a$ | [0.10, 0.38] |
| | Flow index | $n$ | [0.20, 1.00] |
| | Normalized yield aperture | $\log_{10}\beta_y$ | [−2, 0.15] |

Hydraulic outputs were used to isolate rheology-induced changes in the flux-weighted aperture measure before coupling to fracture–matrix heat exchange. For Ellis fluids, the shear-thinning branch contribution is $q_{ST}(a) = C_{PL}a^{2+1/n}$. Since the Ellis active set is the full aperture support, the mean shear-thinning-to-Newtonian flux ratio is $R_E = \langle q_{ST}(a)\rangle_{\mathcal{A}} / \langle q_N(a)\rangle_{\mathcal{A}}$, with corresponding shear-thinning flux fraction $F_{ST} = R_E/(1 + R_E)$. The Ellis hydraulic outputs were

$$H_{E,1} = \langle z\rangle_{w_E} - \langle z\rangle_{w_N}, \qquad H_{E,2} = \ln R_E. \tag{6.9}$$

Here $z(a) = \log_{10}(a/\langle a\rangle_p)$, while $\langle\cdot\rangle_{w_E}$ and $\langle\cdot\rangle_{w_N}$ denote flux-weighted averages computed with the Ellis and Newtonian flux laws, respectively. For lognormal aperture statistics, the centroid shift reduces to

$$H_{E,1} = \frac{\sigma^2}{\ln 10}\left(\beta_{PL} - 3\right) F_{ST}. \tag{6.10}$$

For Herschel–Bulkley fluids, the hydraulic outputs were defined relative to the corresponding power-law fluid with the same $m$ and $n$ and zero yield stress:

$$H_{HB,1} = \ln\left(\frac{\langle q_{HB}(a)\rangle_p}{\langle q_{PL}(a)\rangle_p}\right), \qquad H_{HB,2} = \ln\left(\frac{P_{\mathrm{act}}}{1 - P_{\mathrm{act}}}\right). \tag{6.11}$$

Thus, $H_{HB,1}$ measures the total flow retained relative to the corresponding yield-free power-law reference, whereas $H_{HB,2}$ measures the active geometric aperture fraction on a logit scale.

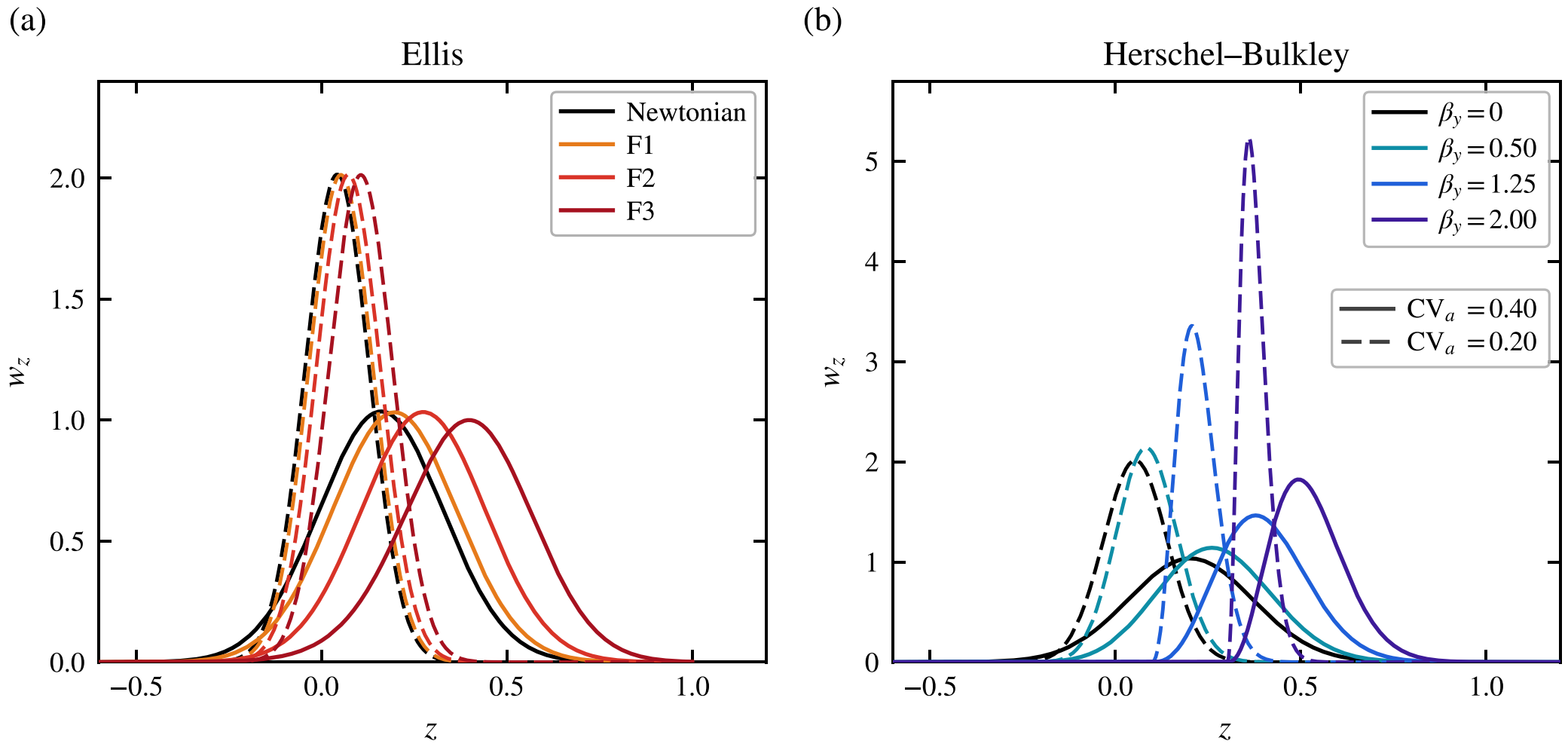

**Fig. 3.** Flux-weighted log-aperture distributions for the deterministic cases, with $z = \log_{10}(a/\langle a\rangle_p)$, $w_z = a\,w(a)$, and $\ln(10)\int w_z\,\mathrm{d}z = 1$. Solid and dashed curves denote $\mathrm{CV}_a = 0.40$ and 0.20. (a) Ellis cases F1–F3 of Table 1, including the Newtonian limit. (b) Herschel–Bulkley cases with $\beta_y = 0$ (power-law reference), 0.50, 1.25, and 2.00. For $\beta_y = 0.50, 1.25, 2.00$, the active probability masses are respectively $1.00, 0.11, 1.6\times10^{-4}$ at $\mathrm{CV}_a = 0.20$, and $0.95, 0.22, 2.3\times10^{-2}$ at $\mathrm{CV}_a = 0.40$. The $\beta_y = 2.00$ curves are the limiting case of Section 6.1.

Jansen estimators used $N = 8192$ samples per base matrix and sixteen independently scrambled Sobol replicates. Reported $S_1$ and $S_T$ values are replicate means, with standard-deviation error bars; $S_T - S_1$ measures interaction effects. Small negative $S_1$ estimates were treated as finite-sample noise.

## 7. Thermal signatures and transport mechanisms

The two constitutive columns isolate different mechanisms and therefore use different null cases. Ellis fluids are compared with their Newtonian limit, whereas Herschel–Bulkley fluids vary the yield threshold at fixed $m$ and $n$ and use the corresponding zero-yield power-law fluid as reference. A Newtonian reference for the latter would confound yield and shear-thinning effects. Each rheology is also compared with its homogeneous planar-channel solution in Fig. 2. Common to Figs. 3–6: the pressure gradient is set in every case to hold $Pe_{\mathrm{th}} = 100$, time is normalized by $t_c$, solid and dashed curves are $\mathrm{CV}_a = 0.40$ and 0.20, and the remaining parameters are those of Table 1.

### *7.1. Hydraulic channel selection*

Fig. 3 shows hydraulic selection in aperture space. Heat samples the flux-weighted density $w(a)$, rather than the geometric density $p(a)$, so the plotted distributions identify the aperture classes carrying the injected thermal power.

For Ellis fluids, the full aperture support remains active. Increasing the shear-thinning contribution shifts the flux-weighted density toward larger apertures, without introducing a hydraulic cutoff. For Herschel–Bulkley fluids, selection is sharper: apertures below the yield aperture are removed from the active set, and flux is further suppressed near the yield edge by $\mathcal{F}_Y(a_y/a)$. The displacement and truncation visible in Fig. 3 are therefore the two basic aperture-space signatures of rheology-induced hydraulic selection.

### *7.2. Thermal breakthrough curves*

Fig. 4 shows the outlet temperature deficit $1-\theta_{\mathrm{out}}$. At fixed $\langle\bar{u}(a)\rangle_w$, differences reflect flux partitioning and matrix exchange; the crossover follows the local memory-to-advection ratio in Eq. (6.3). Since $\lambda(a)^2 \propto L^2/q(a)^2$, shorter or more transmissive channels reach the matrix-controlled regime earlier in absolute time, so the normalized abscissa spans configurations rather than a single observation window. The far-right range exposes the limiting amplitudes, whereas operational observations sample the rheology-dependent crossover.

For Ellis fluids, increasing shear thinning raises the compensated late-time plateau and prolongs the tail by shifting heat-carrying flux toward large, weakly exchanging apertures. The effect strengthens as the aperture distribution broadens.

The Herschel–Bulkley curves remain more compact. Active-set truncation and near-threshold flux suppression act in opposite directions, producing a non-monotone response to $\beta_y$ at fixed flux-weighted mean velocity.

### *7.3. Mean longitudinal displacement*

Fig. 5 shows the normalized first moment $\mathcal{M}(t)/L$ in the semi-infinite reference problem. It complements outlet breakthrough and may exceed unity once the front passes the monitoring section.

For Ellis fluids, shear thinning raises the late-time displacement by shifting heat-carrying flux toward larger apertures; the effect is strongest for the broad distribution. Accordingly, $\mathcal{M}_E/\mathcal{M}_N$ approaches a constant prefactor ratio at late times.

For Herschel–Bulkley fluids, yielding removes small apertures and suppresses near-threshold flux, biasing the retained measure toward mobile channels. The resulting increase is more moderate, and $\mathcal{M}_{HB}/\mathcal{M}_{PL}$ likewise tends to a constant.

### *7.4. Longitudinal spreading*

Fig. 6 shows the longitudinal variance normalized by $L^2$. It combines channel-scale matrix-memory spreading with persistent differences in front position among aperture classes.

For Ellis fluids, spreading is amplified more strongly than mean displacement because the variance samples higher flux moments and is more sensitive to the high-aperture tail, especially for the broader distribution.

For Herschel–Bulkley fluids, variance also increases but remains more compact: spreading grows through velocity contrasts within the active population rather than through an enhanced high-aperture tail.

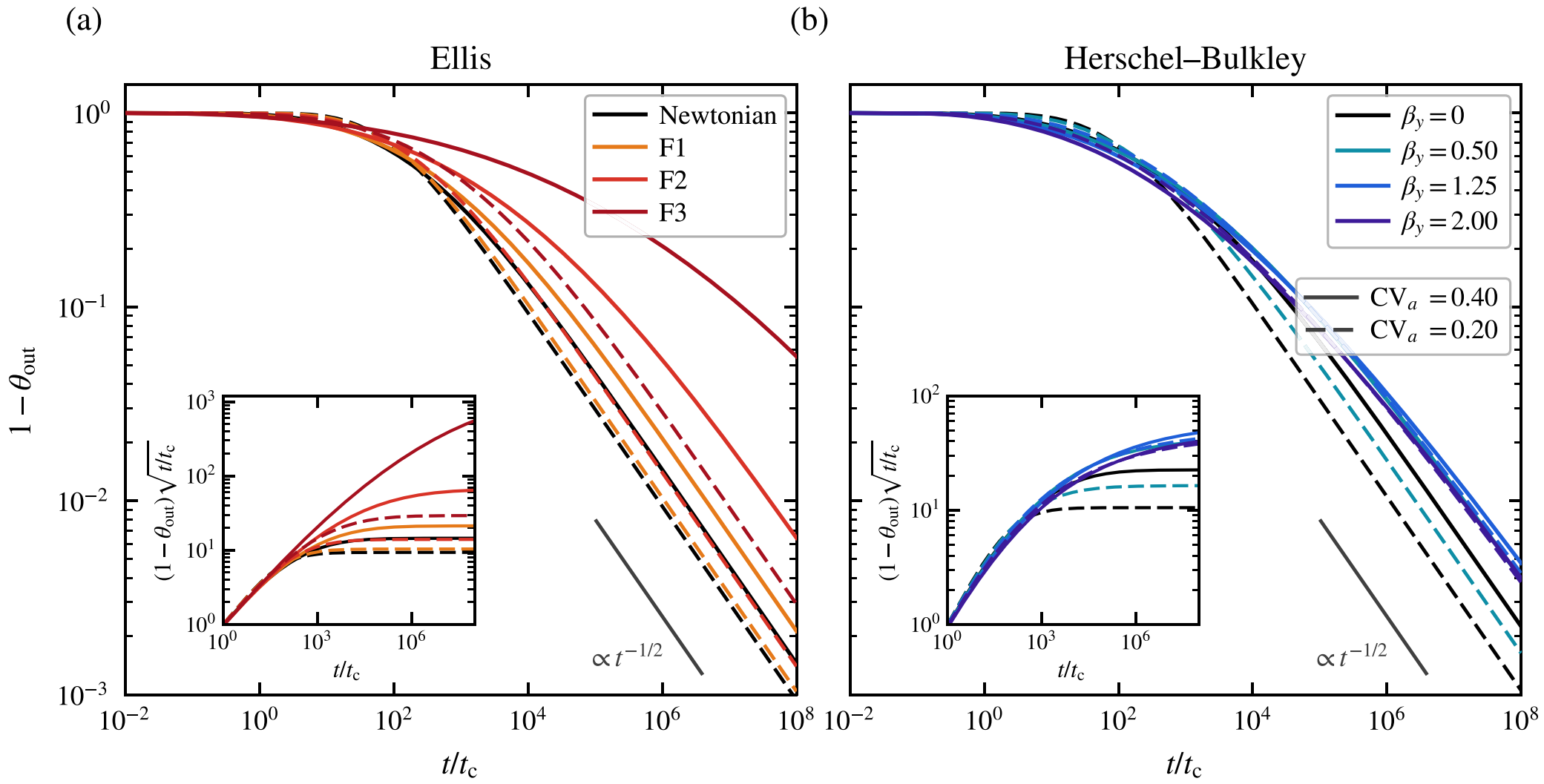

**Fig. 4.** Outlet temperature deficit for (a) Ellis fluids and (b) Herschel–Bulkley fluids in the flux-weighted parallel-channel model; normalization and line styles are defined at the beginning of Section 7. The Ellis panel includes the Newtonian limit and the Herschel–Bulkley panel the corresponding zero-yield power-law reference. Insets show $(1-\theta_{out})\sqrt{t/t_c}$; the plateau confirms the common late-time scaling $1-\theta_{out} \propto t^{-1/2}$. The time axis stops at $t/t_c = 10^8$, by which every compensated curve has reached its plateau; for the reference $t_c$, this upper limit is approximately $3.8 \times 10^2$ yr and is included solely to display the asymptotic amplitude. The Herschel–Bulkley family is not monotone in $\beta_y$: the late-time amplitude is largest at $\beta_y = 1.25$ and smaller again at $\beta_y = 2.00$, so the curves are identified individually rather than by an ordering arrow.

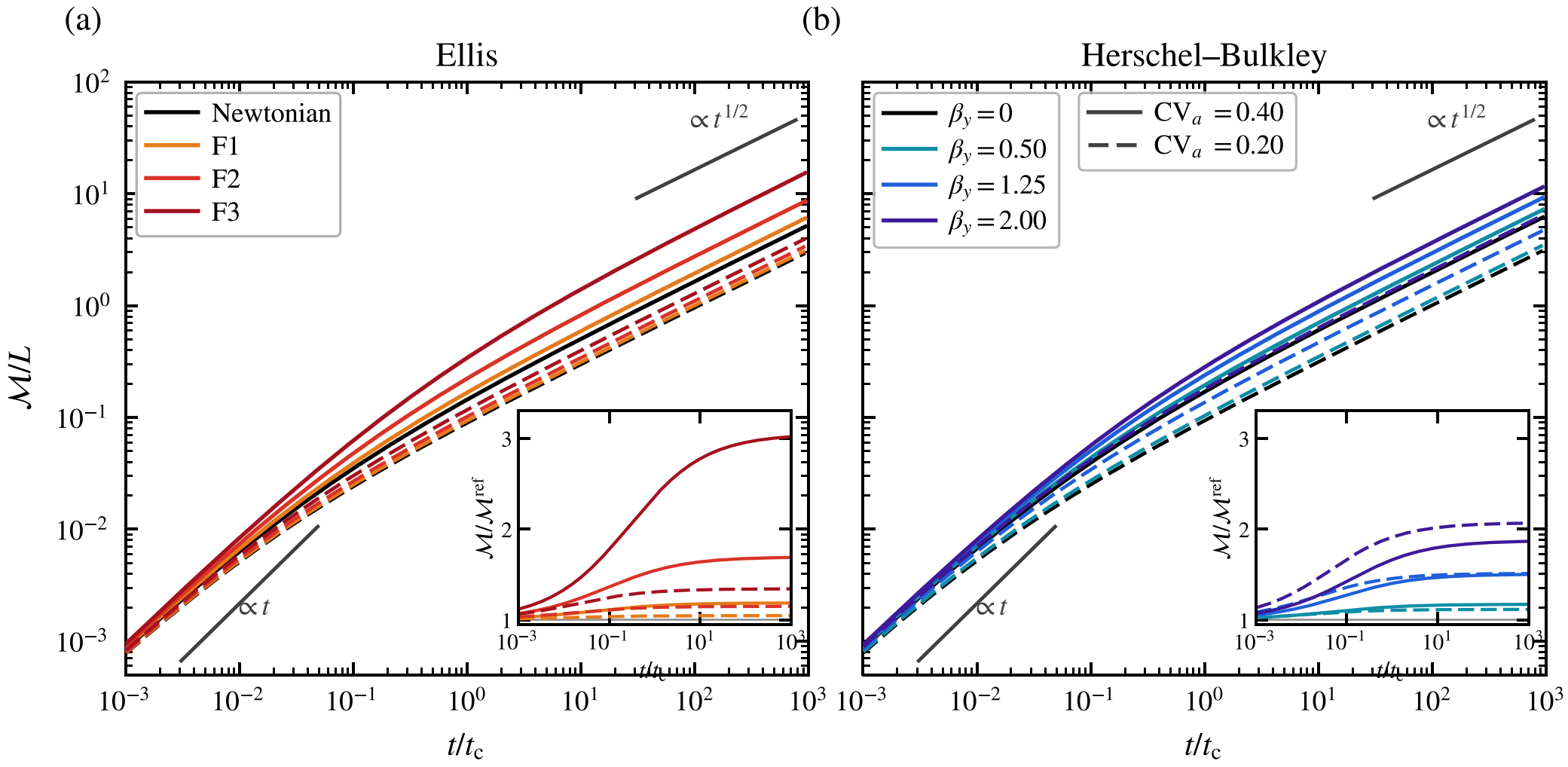

**Fig. 5.** Normalized mean longitudinal displacement $\mathcal{M}(t)/L$ for (a) Ellis fluids and (b) Herschel–Bulkley fluids in the flux-weighted parallel-channel model with lognormal aperture statistics; normalization and line styles as stated at the beginning of Section 7. Reference slopes indicate the transition from early-time ballistic advance, $\mathcal{M} \propto t$, to late-time matrix-controlled growth, $\mathcal{M} \propto t^{1/2}$. The insets show each case referred to the null case of its own family, $\mathcal{M}/\mathcal{M}^{ref}$, where $\mathcal{M}^{ref}$ is the Newtonian limit for the Ellis column and the zero-yield power-law reference for the Herschel–Bulkley column; the ratio approaches a constant once the matrix-controlled regime is reached.

### 7.5. *Hydraulic versus transport sensitivities*

Fig. 7 separates hydraulic sensitivity at fixed driving from transport sensitivity after normalization to the same flux-weighted mean velocity. Total-effect indices include interactions and therefore need not sum to unity.

For Ellis fluids, $H_{E,2} = \ln R_E$ is controlled almost entirely by $\chi_E$. The aperture-shift metric $H_{E,1}$ instead ranks $n$ first, $CV_a$ second, and $\chi_E$ third. This follows directly from Eq. (6.10), where the shift is proportional to $\sigma^2[(1-n)/n]F_{ST}$.

After velocity normalization, Ellis transport is governed mainly by $CV_a$ and $n$; their effects on the breakthrough prefactor are comparable, while $\chi_E$ is secondary. This agrees with the reduced-amplitude similarity of Section 5, although the dimensional prefactors retain additional $\beta_{PL}$-dependent factors and do not collapse exactly.

For Herschel–Bulkley fluids, $\beta_y$ dominates both hydraulic diagnostics, with $CV_a$ secondary. Active support is independent of $n$ because the yielded set is defined by $a > a_y$; retained flow depends weakly on $n$ through the power-law exponent and yield-suppression factor.

For normalized Herschel–Bulkley transport, the breakthrough prefactor ranks $n > \beta_y > CV_a$. The yield parameter also affects the short-time variance coefficient and $F_{inter}$, but is weak for the late-time mean and variance prefactors. Yielding therefore acts chiefly through outlet tailing and the partition between matrix-memory and persistent inter-channel spreading.

For both rheologies, the variance components are controlled mainly by $CV_a$, with $n$ secondary and the transition or yield parameter weak; $F_{inter}$ is therefore the most direct diagnostic of persistent aperture-controlled channel contrast.

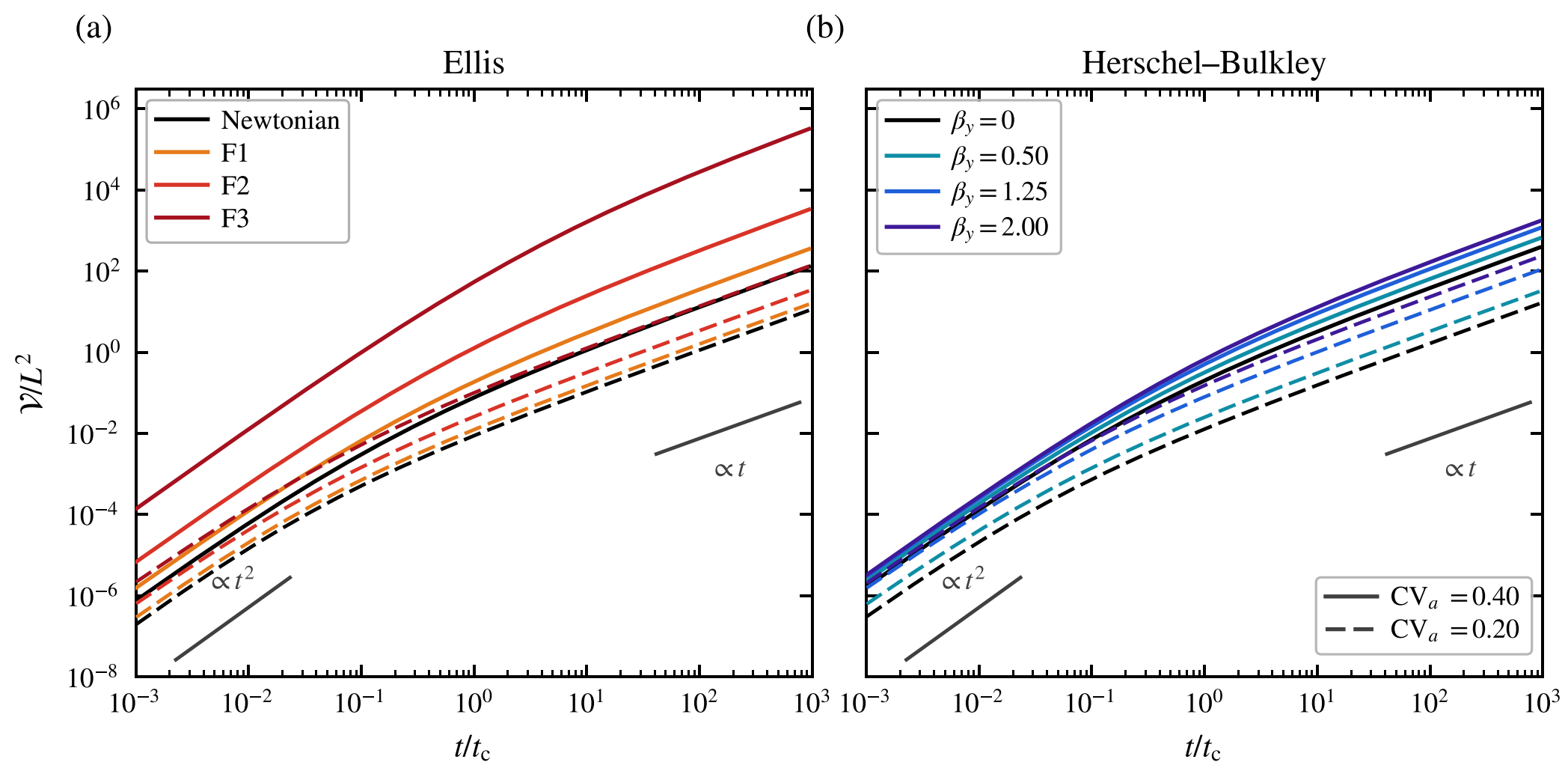

**Fig. 6.** Longitudinal variance $\mathcal{V}/L^2$ for (a) Ellis fluids and (b) Herschel–Bulkley fluids in the flux-weighted parallel-channel model with lognormal aperture statistics; normalization and line styles as stated at the beginning of Section 7. Reference slopes show the formal reduced-model limits $\mathcal{V} \propto t^2$ and $\mathcal{V} \propto t$. The former lies below the aperture-homogenization time for the reference parameters; spreading is interpreted only for $t \gtrsim t_\perp$ as stated in Section 3.1.

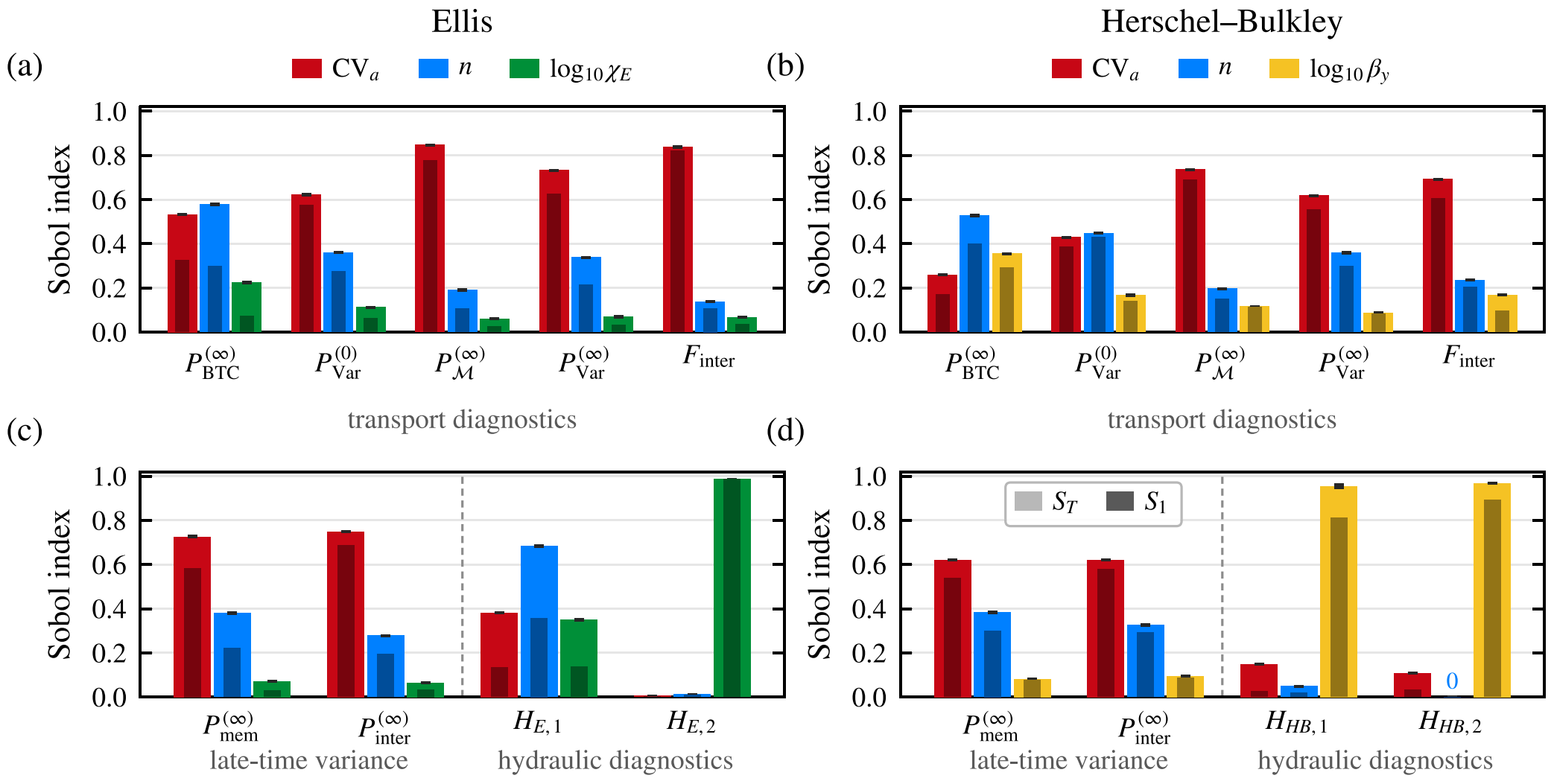

**Fig. 7.** First-order and total-effect Sobol indices for Ellis fluids (left column) and Herschel–Bulkley fluids (right column). Input parameters are $CV_a$, $n$, and either $\log_{10}\chi_E$ or $\log_{10}\beta_y$, over the ranges of Table 3. Panels (a,b) show normalized transport outputs; panels (c,d) show late-time variance components to the left of the rule and unscaled hydraulic diagnostics to its right. Red and blue bars denote $CV_a$ and $n$, green bars $\log_{10}\chi_E$, and yellow bars $\log_{10}\beta_y$. Error bars are standard deviations over scrambled Sobol replicates. Each light bar is a total-effect index and the nested dark bar is $S_1$; their gap measures interactions. The total effect of $n$ on $H_{HB,2}$ is exactly zero and is annotated to distinguish it from a missing bar.

### *7.6. Numerical assessment of the independent-channel limit*

The parallel-channel model is a maximum-persistence reference, not a substitute for a connected two-dimensional fracture model. Flux-weighted superposition isolates aperture variability, nonlinear aperture-to-flux mapping, and fracture–matrix exchange while omitting transverse hydraulic communication and streamline reorganization.

Fig. 8 verifies the formulation where the independent-channel assumption is exact. TDRW simulations use the same longitudinally uniform aperture classes and flux partitioning, and agree in mean position, variance, and breakthrough for the tested Newtonian and power-law closures. This verifies the transport superposition, not Ellis or Herschel–Bulkley flow in a connected aperture field.

In the connected rough field of Fig. 9, mean propagation and outlet survival remain close to the channel prediction over the tested interval $0.05 \leq CV_a \leq 0.15$, whereas transverse pressure redistribution reduces longitudinal variance. Thus the model captures the leading mean and outlet signatures for this realization but overestimates spreading.

This connected-field comparison is limited by the weak-heterogeneity assumptions of Appendix B. The wider deterministic and Sobol ranges instead describe the independent-channel end-member and are limited by the moment-retention criterion of Section 3.3; they do not imply connected-field validation. The exact decomposition in Eq. (3.24) separates $V_{mem} = \langle \sigma_x^2 \rangle_w$ from $V_{inter} = \mathrm{Var}_w[\mathcal{M}(t \mid a)]$. At late times the latter follows the flux contrast, and Eq. (B.11) bounds only its connected-field contribution by the Eulerian factor $\Gamma_q$; finite trajectories can screen it further. The plotted reference carries $V_{mem}$ over unchanged as a closure assumption, so it is a screened estimate rather than a bound on total variance or an exact transient correction. It remains above the resolved results over the tested interval.

No connected-field simulation is presented for Ellis or Herschel–Bulkley rheology. Ellis flow nevertheless combines the two verified monomial closures; its local gradient sensitivity, $\partial \ln q_E / \partial \ln G = [C_N a^3 + n^{-1} C_{PL} a^{2+1/n}]/q_E(a)$, is a flux-weighted average of the Newtonian and power-law exponents. Hence its hydraulic screening is bracketed by the tested values 3/8 and $\Gamma_q(1/n)$, although this does not bracket the full thermal response.

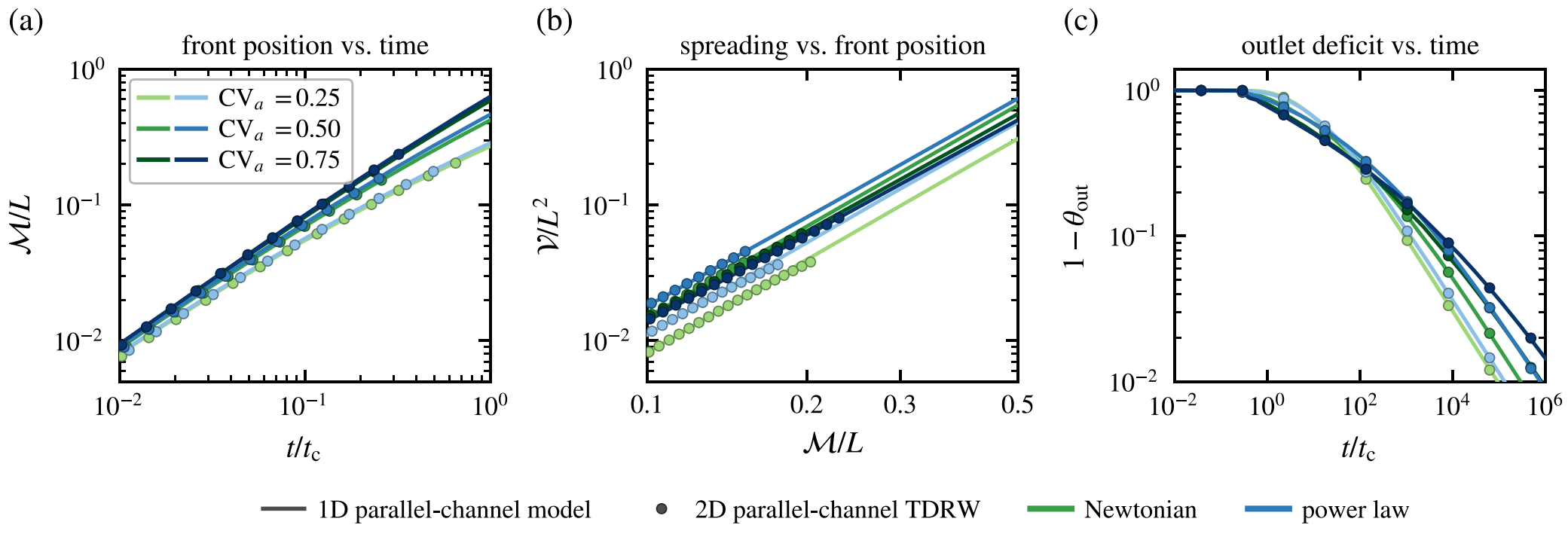

**Fig. 8.** Verification in an exact parallel-channel geometry against two-dimensional TDRW simulations for Newtonian and power-law flow, with $Pe_{\text{th}} = 100$, $N_p = 10^7$, and nominal lognormal $\text{CV}_a = 0.25, 0.50, 0.75$. Both models use the same discretized aperture classes. Green denotes Newtonian flow, blue power-law flow, and darker shades larger $\text{CV}_a$. Panels (a) and (c) show $\mathcal{M}/L$ and $1-\theta_{\text{out}}$ against $t/t_c$; panel (b) shows $\mathcal{V}/L^2$ against $\mathcal{M}/L$, comparing equal front advance. Solid curves are the semi-analytical predictions and symbols the TDRW results. Across the three diagnostics, the root-mean-square relative discrepancy is at most 4.1% and the maximum is below 10%. Variance is restricted to the pre-outlet window $\theta_{\text{out}} < 0.05$ over $10^{-1} \leq \mathcal{M}/L \leq 5 \times 10^{-1}$.

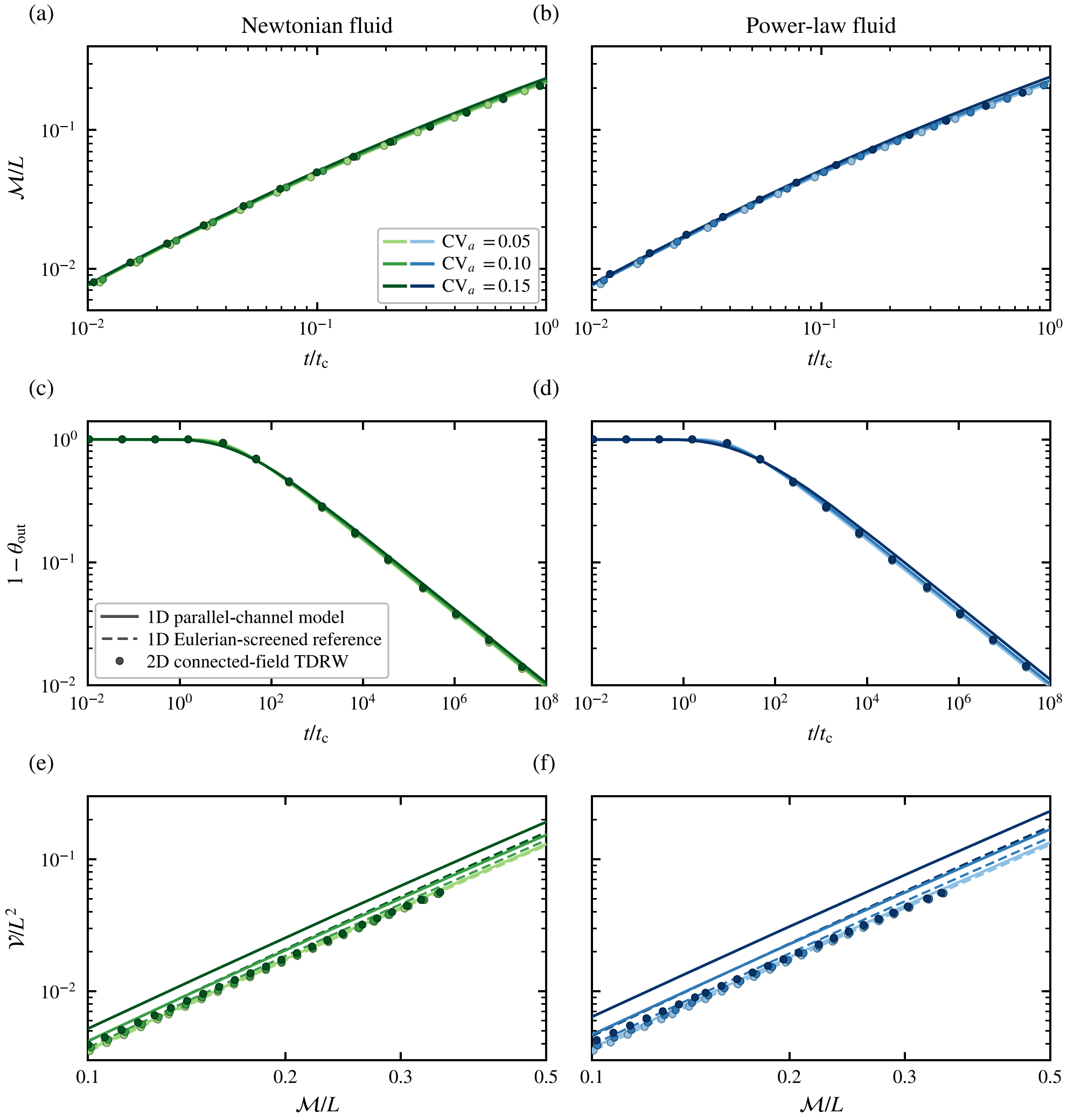

**Fig. 9.** Assessment in heterogeneous connected rough-aperture fields. Semi-analytical parallel-channel predictions are compared with resolved two-dimensional TDRW simulations for Newtonian and power-law flow. The self-affine fields have Hurst exponent 0.8, $L/L_c = 8$ with $L_c = 0.1$ m, $N_x = N_y = 1024$, and $Pe_{\text{th}} = 100$. Empirical $\text{CV}_a = 0.05$, 0.10, and 0.15 are obtained by rescaling one normalized roughness realization, so this is an illustrative realization test, not an ensemble estimate. The columns show Newtonian and power-law flow; panels (a,b), (c,d), and (e,f) show respectively $\mathcal{M}/L$, $1-\theta_{\text{out}}$, and $\mathcal{V}/L^2$. Solid curves are the parallel-channel predictions, filled symbols the connected-field TDRW results, and dashed curves in (e,f) the asymptotic Eulerian-screened references of Appendix B. The dashed curves are neither exact transient corrections nor bounds on total variance. Variance is compared at equal front advance over the pre-outlet window $10^{-1} \leq \mathcal{M}/L \leq 5 \times 10^{-1}$, where $\theta_{\text{out}} < 0.05$.

Herschel–Bulkley flow instead requires a solver that tracks the yield surface and connectivity of the mobile set. Because the present resolved solver is limited to purely viscous shear-thinning flow, the Herschel–Bulkley results are reported only as independent-channel end-member predictions.

## 8. Conclusions

We developed an interpretive, semi-analytical framework linking non-Newtonian rheology to flux-weighted aperture sampling in fracture–matrix heat transport. It targets thermal characterization during engineered injection, not fluid selection for long-term production or replacement of resolved transport models. Within the independent-channel end-member, matrix diffusion fixes the leading late-time exponents, while the aperture-to-flux law controls their amplitudes. Shear thinning increases the hydraulic weight of large apertures, whereas yield stress restricts the active population; variance responds most strongly because it samples higher flux moments than outlet tailing or mean displacement. For monomial flux laws, the reduced late-time amplitudes collapse onto the single similarity variable $g = \beta^2\sigma^2$, so shear thinning acts as an amplified effective aperture variability. Ellis and Herschel–Bulkley flow break this collapse and depend instead on $(\widetilde{\chi}_E, g_N, g_{PL})$ and $(\beta_{PL}, \sigma, \tilde{y})$, respectively. Departures from the homogeneous planar-channel reference grow with aperture variability, Ellis branch dominance, and partial yielding.

The results assume an idealized aperture distribution and omit lateral heat exchange, hydraulic communication, longitudinal conduction, and Taylor dispersion. Late-time signatures are asymptotic, but rheology also controls the crossover sampled over finite observation windows. Sobol rankings separate hydraulic response at fixed driving — controlled by yield or branch dominance — from normalized transport, governed mainly by aperture variability and flow index. These rankings are conditional on the sampled ranges and fixed-mean-velocity normalization. Applications should impose a physically supported aperture range and reassess omitted mechanisms when their time scales become comparable.

TDRW simulations verify the construction for Newtonian and power-law flow. The connected rough-field test covers one rescaled realization over $0.05 \leq \mathrm{CV}_a \leq 0.15$: mean advance and outlet survival remain close to the channel prediction, while connected flow screens persistent flux contrasts and reduces variance. This supports the mechanistic interpretation for that realization, not a statistical low-variability regime. No connected-field validation is claimed for Ellis or Herschel–Bulkley flow or larger variability. Extending the assessment requires a resolved solver that tracks the yield surface and mobile-set connectivity.

## CRediT authorship contribution statement

**Alessandro Lenci:** Writing – review & editing, Writing – original draft, Visualization, Validation, Software, Methodology, Investigation, Funding acquisition, Formal analysis, Data curation, Conceptualization, Project administration. **Irene Daprà:** Writing – review & editing, Supervision.

## Declaration of generative AI and AI-assisted technologies in the manuscript preparation process

During the preparation of this work, the authors used ChatGPT (OpenAI) and Claude (Anthropic) to assist with language refinement, manuscript organization, and consistency checks. After using these tools, the authors reviewed and edited the content as needed and take full responsibility for the content of the published article.

## Funding

A.L. acknowledges funding from the European Union's Horizon Europe research and innovation programme under the Marie Skłodowska-Curie grant agreement No. 101111216, Project GEONEAT — "Complex Fluids in Fractured Geological Media for Enhanced Heat Transfer". Views and opinions expressed are, however, those of the authors only and do not necessarily reflect those of the European Union or the European Research Executive Agency (REA). Neither the European Union nor the granting authority can be held responsible for them. The funding source had no role in the study design, data collection, analysis or interpretation, manuscript preparation, or the decision to submit the article for publication.

## Declaration of competing interest

The authors declare that they have no known competing financial interests or personal relationships that could have appeared to influence the work reported in this paper.

## Appendix A. Constitutive closures and parameter mapping

This appendix summarizes the rheological closures used in the hydraulic model and their parameter identifications. Throughout, $\tau \geq 0$ and $\dot{\gamma} \geq 0$ denote the shear-stress and shear-rate magnitudes, respectively. The main text writes the channel flux laws in terms of a Newtonian viscosity scale $\eta$, defined by $\tau = \eta\dot{\gamma}$, a high-stress consistency scale $m$ and flow index $n$, defined by the power-law closure $\tau = m\dot{\gamma}^n$ (with $n < 1$ for shear thinning), and a yield stress $\tau_y$ for Herschel–Bulkley fluids. Using the same $n$ across the power-law, Ellis, and Herschel–Bulkley closures is a modeling convention that allows the high-stress shear-thinning exponent to be compared across rheologies; it does not imply that the closures represent the same material over the full stress range.

A standard Ellis fluid is written as

$$\dot{\gamma} = \frac{\tau}{\eta_0}\left[1 + \left(\frac{\tau}{\tau_{1/2}}\right)^{\alpha_E - 1}\right], \tag{A.1}$$

where $\eta_0$ is the zero-shear viscosity, $\tau_{1/2}$ is the Ellis transition stress, and $\alpha_E$ controls the high-stress shear-thinning slope; the subscript distinguishes it from the exponent $\alpha = \beta_{PL} - 1$ used in Appendix D. Setting $\alpha_E = 1/n$, the Ellis law separates exactly into a Newtonian contribution and a high-stress power-law contribution,

$$\dot{\gamma} = \frac{\tau}{\eta_0} + \left(\frac{\tau}{m_E}\right)^{1/n}, \qquad m_E = \eta_0^n \tau_{1/2}^{1-n}. \tag{A.2}$$

Here $\eta_0$ is the Ellis zero-shear viscosity and is distinct from the Newtonian viscosity $\eta$ of Table 1, which parameterizes the separate Newtonian case. Identifying the high-stress consistency $m_E$ with the common consistency scale $m$ gives the mapping

$$m = \eta_0^n \tau_{1/2}^{1-n}, \qquad \tau_{1/2} = \left(\frac{m}{\eta_0^n}\right)^{1/(1-n)}, \qquad n \neq 1. \tag{A.3}$$

This mapping is only a parameter convention. The Newtonian reference used in the paper is obtained by suppressing the shear-thinning contribution, not by taking the formal limit $n \to 1$ of the full Ellis relation, because that limit does not recover a Newtonian fluid of viscosity $\eta_0$.

The Herschel–Bulkley closure, $\dot{\gamma} = 0$ for $\tau \leq \tau_y$ and $\dot{\gamma} = [(\tau - \tau_y)/m]^{1/n}$ for $\tau > \tau_y$, uses the same $m$ and $n$ as the power-law closure: it differs from the power-law model only through the yield stress and reduces to it in the limit $\tau_y \to 0$. Accordingly, $\eta$ controls the finite-viscosity Newtonian branch, $m$ and $n$ control the high-stress shear-thinning branch, and $\tau_y$ controls yield-stress activation.

### A.1. Dimensional realization of the deterministic cases

The Ellis parameters $(\eta_0, \tau_{1/2}, n)$ of Table 1 are mapped to the additive Newtonian and power-law coefficients through Eq. (A.3). For each deterministic aperture distribution and rheology, the pressure-gradient magnitude is then adjusted to hold $Pe_{\text{th}}$ fixed; the corresponding pressure drop and wall and yield stresses follow from the relations stated in the caption of Table 2.

This construction differs from the maps of Section 3.3, where the yield stress is a prescribed fluid property and the pressure gradient is swept. In the deterministic set, $\beta_y$ is prescribed and the dimensional yield stress follows from the required gradient. The power-law and Herschel–Bulkley consistency $m = 0.16$ Pa s$^n$ at $n = 0.61$ represents a dilute polymeric carrier (apparent viscosity $\approx 1.4 \times 10^{-2}$ Pa s at $5 \times 10^2$ s$^{-1}$), giving the loadings listed in the table. Ellis F1 instead represents a laboratory silicone oil and requires a gradient beyond field conditions at the prescribed $Pe_{\text{th}}$; it is included only to span the weakly shear-thinning end of the constitutive range.

## Appendix B. Hydraulic screening of the independent-channel limit

This appendix derives the Eulerian screening factor used to correct the persistent inter-channel contribution to the displacement variance in the two-dimensional validation. The derivation is restricted to the monomial hydraulic closures used for the Newtonian and power-law comparisons. In the independent-channel limit, aperture variations are longitudinally persistent and all channels experience the same macroscopic pressure gradient. In a connected two-dimensional fracture, transverse pressure redistribution screens the longitudinal flux contrasts generated by aperture variability.

For the monomial closures, the two-dimensional local closure is written as

$$\mathbf{q} = -K\, a^{\beta} |\nabla \mathcal{P}|^{\gamma - 1} \nabla \mathcal{P}, \tag{B.1}$$

where $\gamma$ is the hydraulic-gradient exponent and $K$ is independent of the hydraulic-gradient magnitude. The Newtonian case has $\beta = 3$, $\gamma = 1$, and

$$K = K_N = \frac{1}{12\eta}.$$

The power-law case has $\beta = \beta_{PL} = 2 + 1/n$, $\gamma = 1/n$, and

$$K = K_{PL} = \frac{n}{2n+1} \left(\frac{1}{2}\right)^{1+1/n} m^{-1/n}.$$

For a uniform hydraulic-gradient magnitude $G_0$, these mobilities recover the main-text prefactors $C_N = K_N G_0$ and $C_{PL} = K_{PL} G_0^{1/n}$ in Eqs. (2.10) and (2.13), respectively. The mobility prefactor $K$ cancels from the normalized screening factor derived below.

Let the aperture field be a weak perturbation about its mean,

$$a(\mathbf{x}) = a_0 \left[1 + \varepsilon h(\mathbf{x})\right], \qquad \langle h \rangle = 0, \qquad \varepsilon \ll 1. \tag{B.2}$$

Let the mean pressure field impose a uniform hydraulic-gradient magnitude $G_0$ in the $x_1$ direction,

$$\nabla \mathcal{P}_0 = -G_0 \mathbf{i}_1, \qquad G_0 > 0. \tag{B.3}$$

The corresponding mean longitudinal flux is

$$q_{\text{ref}} = K a_0^{\beta} G_0^{\gamma}. \tag{B.4}$$

Write the longitudinal flux component as

$$q_1(\mathbf{x}) = q_{\text{ref}} + \delta q_1(\mathbf{x}) + O(\varepsilon^2), \tag{B.5}$$

where $\delta q_1$ is the first-order perturbation induced by the aperture fluctuation in Eq. (B.2). With hats denoting Fourier transforms in the fracture plane and $\mathbf{k} = (k_1, k_2)$ denoting the two-dimensional wavevector, linearization of Eq. (B.1) gives

$$\frac{\widehat{\delta q_1}(\mathbf{k})}{q_{\text{ref}}} = \mathcal{F}_q(\mathbf{k})\, \varepsilon \hat{h}(\mathbf{k}), \qquad \mathcal{F}_q(\mathbf{k}) = \beta \frac{k_2^2}{\gamma k_1^2 + k_2^2}. \tag{B.6}$$

Here $\mathcal{F}_q(\mathbf{k})$ is the dimensionless hydraulic filter that maps aperture perturbations to longitudinal-flux perturbations at first order. The independent-channel limit is recovered for purely transverse aperture modes, $k_1 = 0$, for which $\mathcal{F}_q = \beta$. Longitudinally persistent aperture contrasts are therefore unscreened in the channel model.

For an aperture spectrum $S_h(\mathbf{k}) = |\hat{h}(\mathbf{k})|^2$, define the normalized retained fraction of longitudinal flux variance as

$$\Gamma_q = \frac{\displaystyle\int \mathcal{F}_q^2(\mathbf{k}) S_h(\mathbf{k})\, \mathrm{d}\mathbf{k}}{\displaystyle\beta^2 \int S_h(\mathbf{k})\, \mathrm{d}\mathbf{k}}. \tag{B.7}$$

The normalization is the independent-channel value. Therefore $0 \le \Gamma_q \le 1$, with $\Gamma_q = 1$ corresponding to the unscreened parallel-channel limit.

For a statistically isotropic aperture spectrum, with $k_1 = k\cos\vartheta$ and $k_2 = k \sin\vartheta$, where $\vartheta$ is the wavevector angle, Eq. (B.7) reduces to the angular average

$$\Gamma_q(\gamma) = \left\langle \left[\frac{\sin^2\vartheta}{\gamma\cos^2\vartheta + \sin^2\vartheta}\right]^2 \right\rangle_{\vartheta} = \frac{\sqrt{\gamma} + 2}{2\left(\sqrt{\gamma} + 1\right)^2}. \tag{B.8}$$

Thus, the aperture exponent $\beta$ sets the unscreened monomial aperture-to-flux contrast, while the gradient exponent $\gamma$ controls the fraction of that contrast retained after transverse pressure redistribution.

For Newtonian flow,

$$\Gamma_q^N = \Gamma_q(1) = \frac{3}{8}. \tag{B.9}$$

For power-law flow,

$$\Gamma_q^{PL}(n) = \Gamma_q(1/n) = \frac{\sqrt{n}\,(1 + 2\sqrt{n})}{2(1 + \sqrt{n})^2}. \tag{B.10}$$

For shear-thinning fluids, $n < 1$, this value is smaller than 3/8, corresponding to stronger screening of longitudinal flux contrasts in the weak-heterogeneity isotropic limit.

At late times, Eq. (3.26) gives

$$\mathcal{M}(t \mid a) \sim \frac{2}{\sqrt{\pi}} X(a) t^{1/2}, \qquad X(a) = \frac{q(a)}{2\phi_m \sqrt{D_m}},$$

and hence

$$V_{\text{inter}}(t) \sim \frac{4}{\pi} \mathrm{Var}_w[X(a)]\, t.$$

The first-order perturbation of $X$ is therefore the first-order flux perturbation derived above. Because the inter-channel variance is already of order $\varepsilon^2$, the perturbation of the flux-weighted measure affects this relation only at higher order.

Two distinct mechanisms screen the persistent inter-channel contribution in a connected field, and $\Gamma_q$ accounts for one of them. The first is the transverse pressure redistribution derived above, which acts pointwise on the longitudinal flux and is retained exactly by $\Gamma_q$. The second is Lagrangian: $V_{\text{inter}}$ is the variance of a front position accumulated along a trajectory, so that a path of finite length averages over the aperture fluctuations it traverses. At the same order in $\varepsilon$ this introduces a streamwise filter which, being bounded above by unity, can only reduce the retained variance further, and which becomes more severe as the traveled distance grows relative to the aperture correlation length. It is not evaluated here, because doing so would require the aperture spectrum of the connected field as an additional input, which the independent-channel formulation does not otherwise use. Its sign, however, is fixed. For weak heterogeneity and statistical isotropy, the Eulerian retention factor therefore bounds the connected-field persistent contribution from above,

$$V^{\mathrm{conn}}_{\mathrm{inter}}(t) \le \Gamma_q V^{\mathrm{IC}}_{\mathrm{inter}}(t), \qquad t \to \infty, \tag{B.11}$$

with equality only for longitudinally persistent aperture modes, for which the streamwise filter is inactive. Here $V^{\mathrm{IC}}_{\mathrm{inter}}(t) = \mathrm{Var}_w[\mathcal{M}(t \mid a)]$. The plotted screened reference is defined separately as

$$V_{\mathrm{scr}}(t) \equiv V^{\mathrm{IC}}_{\mathrm{mem}}(t) + \Gamma_q V^{\mathrm{IC}}_{\mathrm{inter}}(t), \qquad V^{\mathrm{IC}}_{\mathrm{mem}}(t) = \langle \sigma_x^2(a,t)\rangle_w. \tag{B.12}$$

Carrying the matrix-memory term over unchanged is a closure assumption, so Eq. (B.12) is neither a bound on total displacement variance nor an exact transient correction. At finite times, $\mathcal{M}(t \mid a)$ is not proportional to $q(a)$; in the short-time limit it is proportional to $\bar{u}(a)$. Consequently, $\Gamma_q$ is an asymptotic late-time factor, and is asserted neither as an exact transient screening coefficient nor as a complete account of the screening at late times.

## Appendix C. Two-dimensional TDRW reference solver

The resolved two-dimensional simulations use the time-domain random walk formulation developed and validated by Lenci et al. (2026b). The hydraulic field was obtained with the generalized Reynolds-equation solver of Lenci et al. (2022b), which provides finite-volume edge velocities on the same mesh used for transport. We summarize only the particle update; the full derivation and validation are given by Lenci et al. (2026b). In this study the resolved solver is used only for the Newtonian and power-law comparisons; connected-field Ellis and Herschel–Bulkley simulations are not included.

The fracture plane is discretized into finite volumes indexed by $j$, with cell aperture $a_j$, cell center $\mathbf{x}_j$, and nearest-neighbor set $\mathcal{N}(j)$. For neighboring cells $j$ and $i$, let $u_{ji}$ be the normal edge velocity across their common face, positive from $j$ to $i$. On the uniform Cartesian grid of spacing $\Delta$, the outward advective transition rate is

$$b_{ji} = \frac{\max(u_{ji}, 0)}{\Delta}, \qquad i \in \mathcal{N}(j). \tag{C.1}$$

The corresponding transition probability and mean mobile residence time are

$$P_{ji} = \frac{b_{ji}}{\sum\limits_{k\in\mathcal{N}(j)} b_{jk}}, \qquad \bar{t}^{f}_{j} = \left(\sum_{k\in\mathcal{N}(j)} b_{jk}\right)^{-1}. \tag{C.2}$$

At each mobile step, the destination cell is sampled from $P_{ji}$ and the mobile residence time $\delta t^f_j$ is sampled from an exponential distribution with mean $\bar{t}^f_j$.

Fracture–matrix heat exchange is represented by a trapping time associated with semi-infinite conductive diffusion into the matrix. In cell $j$, the local exchange coefficient is

$$\kappa_j = \frac{2\phi_m\sqrt{D_m}}{a_j}, \tag{C.3}$$

where $D_m$ is the saturated-matrix thermal diffusivity and $\phi_m$ is the matrix-to-fracture-fluid volumetric heat-capacity ratio. Conditional on the mobile residence time, define

$$\zeta_j = \kappa_j\, \delta t^f_j. \tag{C.4}$$

The associated matrix trapping time is sampled from the one-sided Lévy–Smirnov law as

$$\delta t^m_j = \left[\frac{\zeta_j}{2\,\mathrm{erfc}^{-1}(r)}\right]^2, \qquad r \sim U(0,1). \tag{C.5}$$

This law has Laplace transform $\exp[-\zeta_j\sqrt{s}]$, with $s$ the Laplace variable, and therefore reproduces the semi-infinite-matrix memory kernel used in the advection–conduction model.

If a particle is in cell $j$ at step $\ell$, a destination cell $i \in \mathcal{N}(j)$ is sampled and the particle is advanced according to (the step index $\ell$ is distinct from the flow index $n$)

$$\mathbf{x}^{(\ell+1)} = \mathbf{x}_i, \qquad t^{(\ell+1)} = t^{(\ell)} + \delta t^f_j + \delta t^m_j. \tag{C.6}$$

Particles are released at the inlet cross-section with a probability proportional to the local volumetric flux, so that the ensemble represents the same flow-weighted injection as the semi-analytical construction of Section 3.2; the lateral boundaries are periodic. The procedure is repeated until the particle reaches the outlet boundary or the prescribed maximum simulation time. Particles that reach the outlet are absorbed and recorded in the breakthrough curve, and are excluded from the longitudinal front moments from that time onwards; this is the reason the moment comparison is restricted to the pre-outlet window defined in Section 7.6. Outlet breakthrough curves and longitudinal front moments are obtained by ensemble averaging over independent particle realizations.

All stochastic simulations used $N_p = 10^7$ particles and were performed in MATLAB R2026a. The Mersenne Twister generator was initialized with seed 1 at the beginning of each simulation set.

## Appendix D. Supporting analytical derivations

This appendix collects algebra omitted from the main text to keep the model development focused on the governing response and the flux moments that enter the diagnostics.

### D.1. Reduction of the semi-infinite matrix problem

The normalized matrix temperature is $\theta_m(z_m, t; x_1 \mid a) = [T_m(z_m, t; x_1 \mid a) - T_0]/(T_{\mathrm{in}} - T_0)$, where $T_m$ is the matrix temperature and $z_m$ is the coordinate normal to either channel wall, with $z_m = 0$ at the fluid–matrix interface. The cross-aperture coordinate $x_3$ is centered at the channel mid-plane, so that the two fluid–matrix interfaces are located at $x_3 = \pm a/2$; hence, inside the matrix, $z_m = |x_3| - a/2$ for $|x_3| \ge a/2$, with $z_m$ increasing away from the fracture. For each position $x_1$ along the channel, matrix heat transport is modeled as one-dimensional conduction in $z_m$,

$$\frac{\partial \theta_m}{\partial t} = D_m \frac{\partial^2 \theta_m}{\partial z_m^2}, \qquad D_m = \frac{k_r}{\phi_r \rho_f c_{p,f} + (1-\phi_r)\rho_r c_{p,r}}, \tag{D.1}$$

with $\theta_m(z_m, 0) = 0$, $\theta_m(0, t) = \theta(x_1, t)$, and $\theta_m \to 0$ as $z_m \to \infty$. The matrix-to-fluid heat-capacity ratio is

$$\phi_m = \phi_r + (1-\phi_r)\frac{\rho_r c_{p,r}}{\rho_f c_{p,f}}. \tag{D.2}$$

The two-wall exchange term in the cross-sectionally averaged fracture balance is

$$\mathcal{B} = -\frac{2\phi_m D_m}{a}\left.\frac{\partial \theta_m}{\partial z_m}\right|_{z_m=0}, \qquad \frac{\partial \theta}{\partial t} + \bar{u}(a)\frac{\partial \theta}{\partial x_1} = -\mathcal{B}. \tag{D.3}$$

Laplace transformation gives

$$\hat{\theta}_m = \hat{\theta}\exp\!\left(-z_m\sqrt{s/D_m}\right), \qquad \hat{\mathcal{B}} = \kappa(a)\sqrt{s}\,\hat{\theta}, \tag{D.4}$$

which yields Eq. (3.1) and hence the single-channel solution in Eq. (3.6).

Eq. (3.6) is the classical planar-fracture advection–conduction solution of Lauwerier (1955), written for a channel of aperture $a$ carrying a flux $q(a)$ per unit width. The correspondence is obtained by identifying the advective arrival time $t_a = La/q(a)$ and the matrix-memory parameter $\lambda(a) = 2\phi_m\sqrt{D_m}\,L/q(a)$, in which the factor of two accounts for conduction through both walls, $\phi_m$ for the volumetric heat-capacity contrast between matrix and fluid, and $\sqrt{D_m}$ for the semi-infinite conductive memory. The rheology enters only through $q(a)$, so the same closed form holds for every constitutive law considered here. The identification is exact, so the homogeneous limit of Section 3.3 is recovered analytically rather than numerically: what the maps add there is the departure of the heterogeneous non-Newtonian response from the published classical prediction; they do not define a universal adequacy threshold.

### D.2. Moment inversion and the short-time correction

Substituting Eq. (3.3) into $p_x = -\partial\theta/\partial x_1$ and evaluating $\int_0^\infty x_1^\ell \hat{p}_x \, dx_1$ gives

$$\widehat{\mathcal{M}}(s \mid a) = \frac{\bar{u}(a)}{s[s+\kappa(a)\sqrt{s}]}, \qquad \widehat{\mathcal{M}}_2(s \mid a) = \frac{2\bar{u}(a)^2}{s[s+\kappa(a)\sqrt{s}]^2}. \tag{D.5}$$

Their inverse transforms are Eqs. (D.6) and (D.7):

$$\mathcal{M}(t \mid a) = \frac{\bar{u}(a)}{\kappa(a)^2}\left[e^{\kappa(a)^2 t}\,\mathrm{erfc}\left(\kappa(a)\sqrt{t}\right) - 1 + \frac{2\kappa(a)\sqrt{t}}{\sqrt{\pi}}\right], \tag{D.6}$$

and

$$\mathcal{M}_2(t \mid a) = \frac{2\bar{u}(a)^2}{\kappa(a)^4}\left[\kappa(a)^2 t + 3 + \left(2\kappa(a)^2 t - 3\right)e^{\kappa(a)^2 t} \times \mathrm{erfc}\left(\kappa(a)\sqrt{t}\right) - \frac{6\kappa(a)\sqrt{t}}{\sqrt{\pi}}\right]. \tag{D.7}$$

Expanding these expressions at short times gives

$$\mathcal{M}(t \mid a) = \bar{u}(a)t - \frac{4\bar{u}(a)\kappa(a)}{3\sqrt{\pi}}t^{3/2} + O(t^2), \qquad \sigma_x^2(a,t) \sim \frac{8\bar{u}(a)^2\kappa(a)}{15\sqrt{\pi}}t^{5/2}. \tag{D.8}$$

Consequently,

$$\mathcal{V}(t) = \mathrm{Var}_w[\bar{u}(a)]t^2 + A_{5/2}t^{5/2} + O(t^3), \tag{D.9}$$

where

$$A_{5/2} = \frac{8}{15\sqrt{\pi}}\left\langle \bar{u}(a)^2\kappa(a)\right\rangle_w - \frac{8}{3\sqrt{\pi}}\mathrm{Cov}_w(\bar{u}(a), \bar{u}(a)\kappa(a)). \tag{D.10}$$

### D.3. Rheology-specific lognormal moments

For a full-support monomial flux $q = Ca^\beta$, substitution of Eq. (4.2) into the three master prefactors gives

$$P_{\mathrm{BTC}}^{(\infty)} = \frac{2\phi_m\sqrt{D_m}L}{\sqrt{\pi}C}\exp\left(-\beta\mu - \tfrac{1}{2}\beta^2\sigma^2\right), \tag{D.11}$$

$$P_{\mathcal{M}}^{(\infty)} = \frac{C}{\phi_m\sqrt{\pi D_m}}\exp\left(\beta\mu + \tfrac{3}{2}\beta^2\sigma^2\right), \tag{D.12}$$

$$P_{\mathcal{V}}^{(\infty)} = \frac{C^2}{2\phi_m^2 D_m}\left[\exp\left(2\beta\mu + 4\beta^2\sigma^2\right) - \frac{2}{\pi}\exp\left(2\beta\mu + 3\beta^2\sigma^2\right)\right]. \tag{D.13}$$

The Newtonian and power-law results follow by setting $(C,\beta) = (C_N, 3)$ and $(C_{PL}, \beta_{PL})$, respectively.

For the Ellis closure, the finite sum in Eq. (4.5) becomes

$$\langle q_E(a)^r\rangle_{\mathcal{A}} = \sum_{j=0}^{r}\binom{r}{j}C_N^{r-j}C_{PL}^j\exp\left\{v_{rj}\mu + \tfrac{1}{2}v_{rj}^2\sigma^2\right\}, \quad v_{rj} = 3(r-j) + j\beta_{PL}, \qquad r = 1,2,3. \tag{D.14}$$

Because the Ellis flux is the sum of a Newtonian and a shear-thinning monomial branch, the one-parameter similarity of the monomial reference does not apply. The Ellis response contains two median-aperture hydraulic scales, $C_N\exp(3\mu)$ and $C_{PL}\exp(\beta_{PL}\mu)$, associated with the Newtonian and shear-thinning branches, respectively. Their ratio defines the reduced Ellis branch coordinate of Section 5,

$$\widetilde{\chi}_E = \frac{C_{PL}\exp(\beta_{PL}\mu)}{C_N\exp(3\mu)}, \tag{D.15}$$

and the corresponding monomial similarity factors are

$$g_N = 9\sigma^2, \qquad g_{PL} = \beta_{PL}^2\sigma^2. \tag{D.16}$$

In these variables, the Ellis flux moments of Eq. (D.14) take the scaled form

$$\langle q_E(a)^r\rangle_{\mathcal{A}} = \left[C_N\exp(3\mu)\right]^r m_r(\widetilde{\chi}_E; g_N, g_{PL}), \qquad r = 1,2,3, \tag{D.17}$$

where the dimensionless mixed-moment function is

$$m_r(\widetilde{\chi}_E; g_N, g_{PL}) = \sum_{j=0}^{r}\binom{r}{j}\widetilde{\chi}_E^j\exp\left\{\frac{1}{2}\left[(r-j)\sqrt{g_N} + j\sqrt{g_{PL}}\right]^2\right\}. \tag{D.18}$$

Factoring out the Newtonian median-aperture hydraulic scale and its powers gives the reduced late-time prefactors

$$\widetilde{P}_{\mathrm{BTC}}^{(\infty)} = \frac{1}{m_1}, \qquad \widetilde{P}_{\mathcal{M}}^{(\infty)} = \frac{m_2}{m_1}, \qquad \widetilde{P}_{\mathcal{V}}^{(\infty)} = \frac{m_3}{m_1} - \frac{2}{\pi}\left(\frac{m_2}{m_1}\right)^2, \tag{D.19}$$

so that the reduced Ellis response depends on the three groups $(\widetilde{\chi}_E, g_N, g_{PL})$. The branch ratio evaluated at the arithmetic mean aperture, defined in Eq. (2.18), is related to the median-aperture ratio by

$$\chi_E = \widetilde{\chi}_E\exp\left[\frac{1}{2}(\beta_{PL} - 3)\sigma^2\right]. \tag{D.20}$$

The balance between the pure Newtonian and pure shear-thinning contributions to the $r$th flux moment follows from equating the $j = 0$ and $j = r$ terms of Eq. (D.14):

$$\widetilde{\chi}_{E,c}(r) = \exp\left[-\frac{r}{2}\left(\beta_{PL}^2 - 9\right)\sigma^2\right] = \left[\widetilde{\chi}_{E,c}(1)\right]^r, \tag{D.21}$$

which decreases with the moment order $r$ for shear-thinning fluids ($\beta_{PL} > 3$).

For the Herschel–Bulkley closure, the yield threshold is located within the lognormal aperture population by the standardized coordinate

$$\tilde{y} = \frac{\ln a_y - \mu}{\sigma} = \frac{\ln\beta_y + \sigma^2/2}{\sigma}, \qquad \beta_y = \frac{a_y}{\langle a\rangle_p}, \tag{D.22}$$

so that $P_{\mathrm{act}} = 1 - \Phi(\tilde{y})$, with $Z = (\ln a - \mu)/\sigma$ the standardized log-aperture. Let $\alpha = \beta_{PL} - 1 = 1 + 1/n$. The active flux can be factorized as

$$q_{HB}(a) = C_{PL}(a - a_y)^\alpha\left(a + \frac{a_y}{\alpha}\right), \qquad a > a_y, \tag{D.23}$$

which is equivalent to the yield factor in Section 2.4. Substitution of $a = \exp(\mu + \sigma Z)$ and $a_y = \exp(\mu + \sigma\tilde{y})$ gives the dimensionless quadrature

$$\mathcal{G}_r(\beta_{PL}, \sigma, \tilde{y}) = \int_{\tilde{y}}^{\infty}\left(e^{\sigma Z} - e^{\sigma\tilde{y}}\right)^{r\alpha}\left(e^{\sigma Z} + \frac{1}{\alpha}e^{\sigma\tilde{y}}\right)^r\varphi(Z)\,dZ, \tag{D.24}$$

where $\varphi$ is the standard normal density. The yielded flux moments then take the reduced form

$$\langle q_{HB}(a)^r\rangle_{\mathcal{A}} = \left[C_{PL}\exp(\beta_{PL}\mu)\right]^r\mathcal{G}_r(\beta_{PL}, \sigma, \tilde{y}), \qquad r = 1,2,3, \tag{D.25}$$

and factoring out the power-law median-aperture hydraulic scale and its powers gives the reduced late-time prefactors

$$\widetilde{P}_{\mathrm{BTC}}^{(\infty)} = \frac{P_{\mathrm{act}}}{\mathcal{G}_1}, \qquad \widetilde{P}_{\mathcal{M}}^{(\infty)} = \frac{\mathcal{G}_2}{\mathcal{G}_1}, \qquad \widetilde{P}_{\mathcal{V}}^{(\infty)} = \frac{\mathcal{G}_3}{\mathcal{G}_1} - \frac{2}{\pi}\left(\frac{\mathcal{G}_2}{\mathcal{G}_1}\right)^2, \tag{D.26}$$

controlled by $(\beta_{PL}, \sigma, \tilde{y})$; since the mapping $(\mathrm{CV}_a, \beta_y) \mapsto (\sigma, \tilde{y})$ is one-to-one for $\mathrm{CV}_a > 0$ and $\beta_y > 0$, the physical variables $(\mathrm{CV}_a, n, \beta_y)$ span the same reduced similarity space. As $a_y \to 0$, $P_{\mathrm{act}} \to 1$ and $\langle q_{HB}^r\rangle_{\mathcal{A}} \to C_{PL}^r\langle a^{r\beta_{PL}}\rangle_p$, recovering the power-law limit.

## Data availability

Model, Sobol, figure-generation, HDF5 output, and independent verification scripts are archived at https://doi.org/10.5281/zenodo.21168878, which resolves to the latest version. The Sobol indices in Fig. 7 use the restricted domain of Section Section 6.2, $N = 8192$, and sixteen scrambled replicates. Two-dimensional simulations use the flow solver of Lenci et al. (2022b) and the TDRW solver of Lenci et al. (2026b).